\documentclass[twocolumn]{aastex63}
\usepackage{graphicx}
\usepackage{amssymb}
\usepackage{amsmath}
\usepackage{comment}
\usepackage{color}
\usepackage{natbib} 

\def\lya{Ly$\alpha$}

\begin{document}

 \title{LAGER Ly$\alpha$ Luminosity Function at $z\sim7$, Implications for Reionization}

\noindent \correspondingauthor{Isak G. B. Wold} \email{isak.g.wold@nasa.gov}

\author{Isak G. B. Wold}
\affil{Astrophysics Science Division, NASA Goddard Space Flight Center, 8800 Greenbelt Road, Greenbelt, Maryland, 20771, USA}

\author{Sangeeta Malhotra} 
\affiliation{Astrophysics Science Division, NASA Goddard Space Flight Center, 8800 Greenbelt Road, Greenbelt, Maryland, 20771, USA}

\author{James Rhoads} 
\affiliation{Astrophysics Science Division, NASA Goddard Space Flight Center, 8800 Greenbelt Road, Greenbelt, Maryland, 20771, USA}

\author{Junxian Wang}
\affiliation{CAS Key Laboratory for Research in Galaxies and Cosmology, Department of Astronomy, University of Science and Technology of China, Hefei, Anhui 230026, People\textsc{\char13}s Republic of China}
\affiliation{School of Astronomy and Space Science, University of Science and Technology of China, Hefei 230026, People\textsc{\char13}s Republic of China}

\author{Weida Hu}
\affiliation{CAS Key Laboratory for Research in Galaxies and Cosmology, Department of Astronomy, University of Science and Technology of China, Hefei, Anhui 230026, People\textsc{\char13}s Republic of China}
\affiliation{School of Astronomy and Space Science, University of Science and Technology of China, Hefei 230026, People\textsc{\char13}s Republic of China}

\author{Lucia A. Perez}
\affiliation{School of Earth and Space Exploration, Arizona State University, Tempe, AZ 85287, USA}

\author{Zhen-Ya Zheng}
\affiliation{CAS Key Laboratory for Research in Galaxies and Cosmology, Shanghai Astronomical Observatory, Shanghai 200030, People\textsc{\char13}s Republic of China}

\author{Ali Ahmad Khostovan}
\affiliation{Astrophysics Science Division, NASA Goddard Space Flight Center, 8800 Greenbelt Road, Greenbelt, Maryland, 20771, USA}

\author{Alistair R. Walker}
\affiliation{Cerro Tololo Inter-American Observatory, NSF’s NOIRLab, Casilla 603, La Serena, Chile}

\author{L. Felipe Barrientos}
\affiliation{Instituto de Astrof{\'{\i}}sica and Centro de Astroingenier{\'{\i}}a, Facultad de F{\'{i}}sica, Pontificia Universidad Cat{\'{o}}lica de Chile, Casilla 306, Santiago 22, Chile}   

\author{Jorge Gonz\'alez-L\'opez}
\affiliation{N\'ucleo de Astronom\'ia de la Facultad de Ingenier\'ia y Ciencias, Universidad Diego Portales, Av. Ej\'ercito Libertador 441, Santiago, Chile}
\affiliation{Instituto de Astrof{\'{\i}}sica and Centro de Astroingenier{\'{\i}}a, Facultad de F{\'{i}}sica, Pontificia Universidad Cat{\'{o}}lica de Chile, Casilla 306, Santiago 22, Chile}

\author{Santosh Harish}
\affiliation{School of Earth and Space Exploration, Arizona State University, Tempe, AZ 85287, USA}

\author{Leopoldo Infante}
\affiliation{Las Campanas Observatory, Carnegie Institution of Washington, Casilla 601, La Serena, Chile}

\author{Chunyan Jiang} 
\affiliation{CAS Key Laboratory for Research in Galaxies and Cosmology, Shanghai Astronomical Observatory, Shanghai 200030, People\textsc{\char13}s Republic of China}

\author{John Pharo}
\affiliation{School of Earth and Space Exploration, Arizona State University, Tempe, AZ 85287, USA}

\author{Crist\'obal Moya-Sierralta}
\affiliation{Instituto de Astrof{\'{\i}}sica and Centro de Astroingenier{\'{\i}}a, Facultad de F{\'{i}}sica, Pontificia Universidad Cat{\'{o}}lica de Chile, Casilla 306, Santiago 22, Chile}

\author{Franz E. Bauer}
\affiliation{Instituto de Astrof{\'{\i}}sica and Centro de Astroingenier{\'{\i}}a, Facultad de F{\'{i}}sica, Pontificia Universidad Cat{\'{o}}lica de Chile, Casilla 306, Santiago 22, Chile}   
\affiliation{Millennium Institute of Astrophysics, Nuncio Monse{\~{n}}or S{\'{o}}tero Sanz 100, Of 104, Providencia, Santiago, Chile} 
\affiliation{Space Science Institute, 4750 Walnut Street, Suite 205, Boulder, Colorado 80301}

\author{Gaspar Galaz}
\affiliation{Instituto de Astrof{\'{\i}}sica and Centro de Astroingenier{\'{\i}}a, Facultad de F{\'{i}}sica, Pontificia Universidad Cat{\'{o}}lica de Chile, Casilla 306, Santiago 22, Chile}

\author{Francisco Valdes}
\affiliation{National Optical Astronomy Observatory, 950 N. Cherry Avenue, Tucson, AZ 85719, USA}

\author{Huan Yang}
\affiliation{Las Campanas Observatory, Carnegie Institution of Washington, Casilla 601, La Serena, Chile}
\begin{abstract}
We present a new measurement of the \lya\ luminosity function at redshift $z=6.9$, 
finding moderate evolution from $z=5.7$ that is consistent with a fully or largely 
ionized $z\sim7$ intergalactic medium.  
Our result is based on four fields of the LAGER (Lyman Alpha Galaxies in the Epoch of Reionization)
project.   Our survey volume  of $6.1\times10^{6}$ Mpc$^{3}$ is double that
of the next largest $z\sim 7$ survey.   We combine two new LAGER fields (WIDE12 and
GAMA15A) with two previously reported LAGER fields (COSMOS and CDFS).
In the new fields, we: identify $N=95$ new $z=6.9$ Ly$\alpha$ emitter
(LAEs) candidates;
characterize our survey's completeness and reliability; and compute
Ly$\alpha$ luminosity functions.   The best-fit Schechter luminosity
function parameters for all four LAGER fields are in good general agreement.   Two
fields (COSMOS and WIDE12) show evidence for a bright-end excess above
the Schechter function fit.   We find that the Ly$\alpha$ luminosity density
declines at the same rate as the UV continuum LF from $z=5.7$ to 
$z=6.9$.   This is consistent with an intergalactic medium that was fully ionized as early 
as redshift $z\sim 7$, or with a  volume-averaged neutral hydrogen fraction of
$x_{HI} < 0.33$ at $1\sigma$.
\end{abstract}

\section{Introduction}

Ly$\alpha$ emission is intrinsically one of the most luminous emission
lines in the ionized nebula produced by star-forming galaxies. This
bright feature has enabled observational surveys to efficiently obtain
large samples of Ly$\alpha$ emitting galaxies over a wide redshift
range $z=0-8$ \citep[e.g.,][]{cowie98, hu98, rhoads00, malhotra02, ouchi03, gronwall07, gawiser07, deharveng08, hu10, ouchi10, tilvi20, cowie11, blanc11, finkelstein13,  konno14, matthee15, santos16, konno16, konno18, wold17}.
These studies have revealed the general picture that Ly$\alpha$ emitters
(LAEs) are young starbursting galaxies with low masses, low dust content,
and high excitation states that become more common and more luminous
at high-redshifts. 

Furthermore, Ly$\alpha$ emission is one of the few probes of the
ionization state of the intergalactic medium (IGM) during the reionization
epoch \citep{malhotra04,malhotra06}. We know that reionization should fall within the $6<z<9$ redshift
range from the saturation of Ly$\alpha$ absorbers in $z\sim6$ quasar
spectra \citep{Fan06} and from polarization measurements of the cosmic
microwave background \citep{planck18}. Ly$\alpha$ luminosity functions
and their evolution can be used to further constrain the timing of
reionization because Ly$\alpha$ emission is resonantly scattered
by any neutral hydrogen that it encounters, making it very sensitive
to the ionization state of the IGM. Beyond a redshift of $z\sim6$,
previous studies have found that Ly$\alpha$ LFs decline \citep[e.g.,][]{konno14,inoue18}
at a rate exceeding the decline seen in UV LFs \citep[e.g.,][]{finkelstein15, bouwens15}.
This decline may arise from the increasing opacity of the $z\gtrsim7$
IGM and the onset of the reionization epoch. However, the sample sizes
of LAEs at the highest redshifts are still limited ($N<100$), inhibiting
the current Ly$\alpha$ based reionization constraints from
distinguishing between competing theoretical models
\citep[e.g.,][]{robertson15,finkelstein19,kulkarni19a,naidu20}. Furthermore, compiling narrow-band surveys from the literature to study the evolution
of the Ly$\alpha$ luminosity functions can introduce systematics
caused by differences in the adopted NB excess cut and by differences in the method used
to compute the area masked by foreground sources \citep[see discussion
of selection completeness in:][]{hu19}.  These effects can
mimic redshift evolution if not properly accounted for.

With the Lyman-Alpha Galaxies in the Epoch of Reionization (LAGER)
project, we are conducting a definitive $\sim24$ deg$^{2}$ narrow-band
Ly$\alpha$ survey at $z=6.9$ to precisely measure the timing and
morphology of reionization.  LAGER exploits DECam's unique combination
of a large 4-m aperture and $3$ deg$^{2}$ field of view together
with a high detector sensitivity in the near-infrared. DECam's wide
field of view is needed to mitigate cosmic variance, which is expected
to be amplified in any patchy reionization scenario \citep[e.g.,][]{jensen14}.
We have designed and acquired 
a narrow-band filter with a central wavelength of $9642$ \AA\ that
avoids strong sky OH emission lines and atmospheric absorption \citep{zheng19}.
Consequently, LAGER is an extremely efficient Ly$\alpha$ survey at
the epoch of reionization.

The LAGER collaboration has published our Ly$\alpha$ survey results
from the COSMOS and CDFS fields \citep{zheng17,hu19,hu21}, spectroscopic
followup of eight LAEs found within these fields \citep{hu17,yang19},
and our H$\alpha$, {[}OIII{]}, and {[}OII{]} survey results from
the COSMOS field \citep{khostovan20}.

With the addition of the LAGER fields WIDE12 and GAMA15A, we present
four out of the eight currently planned LAGER fields. Even at this
nominal halfway point, the 4-Field LAGER survey represents the largest
$z\sim7$ Ly$\alpha$ survey to date. In this paper, we carefully
correct for selection effects to study the evolution of the Ly$\alpha$
luminosity functions from $z=5.7$ to $6.9$. We find that the evolution
mirrors the decline seen in the UV LFs, which is consistent with a fully
ionized $z=6.9$ neutral hydrogen fraction.

One of the best options to conduct unbiased
  large-volume surveys at high-redshift is
the use of space-based grism instruments \citep[e.g.,][]{Malhotra05,rhoads09,rhoads13,tilvi16,Larson18}.  These slitless spectroscopic surveys are able to avoid bright
skylines that plague ground-based surveys -- especially at
high-redshift.  The upcoming \textit{Nancy Grace Roman Space Telescope} will
have a wide-field (0.281 square degrees) near-infrared (1-1.93 $\mu$m) grism capability that offers the
opportunity to revolutionize $z>8$ Ly$\alpha$ surveys. The LAGER
project provides a reference $z\sim7$ Ly$\alpha$ survey that can be
used in combination with upcoming $z>8$ \textit{Roman} surveys to
study the evolution of the Ly$\alpha$ population and further constrain the ionization state of the IGM.  

Throughout this work, all Ly$\alpha$ equivalent widths (EWs) are
rest-frame and all magnitudes are in the AB magnitude system ($m_{\mbox{\footnotesize{AB}}}=31.4-2.5 \log_{10}f_{\nu}$
with $f_{\nu}$ in units of nJy). We adopt a flat $\Lambda$CDM cosmology
with $\Omega_{m}=0.3$, $\Omega_{\Lambda}=0.7$, and H$_{0}=70$ km
s$^{-1}$Mpc$^{-1}$.

\section{Observations}

\noindent \begin{deluxetable*}{ccccccc} 
\tablecolumns{7} 
\tablewidth{0pc} 
\tablecaption{Exposure Time, Seeing, and Image Depth}
\tablehead{ 
\colhead{Field} \vspace{-0.2cm} & \colhead{Filter} & \colhead{Exp.\ Time} & \colhead{Seeing} & \colhead{Aperture $5\sigma$ Depth} & \colhead{Total $5\sigma$ Depth}  & \colhead{Depth Aperture}\\
\colhead{ } & \colhead{ } & \colhead{(ks)} & \colhead{(arcsec)} & \colhead{(AB mag)} & \colhead{(AB mag)} & \colhead{(Diameter in arcsec)}}
\startdata 
WIDE12 & $\rm{NB964}$ & 100.6 & 1.10 & 25.3/24.7\phm{adf} & 24.6/24.3\phm{adf} & 1.2/1.8\phm{adf} \\
$12^{\rm{h}}04^{\rm{m}}25^{\rm{s}}$ & $\rm{HSC-}y$ & 1.2 & 0.67 & 24.8/24.4\tablenotemark{\rm{\footnotesize{psf}}} & 24.4/24.1\tablenotemark{\rm{\footnotesize{psf}}} & 1.0/1.8\tablenotemark{\rm{\footnotesize{psf}}} \\
$0^{\circ}00^{\prime}00^{\prime\prime}$ & $\rm{HSC-}griz$ & 3.6 & 0.63 & 27.0/26.1\phm{adf} & 26.7/26.0\phm{adf} & 1.0/1.8\phm{adf} \\
(J2000) & $\rm{HSC-}z$ & 1.2 & 0.68 & 25.9/25.1\phm{adf} & 25.5/25.0\phm{adf} & 1.0/1.8\phm{adf} \\
  & $\rm{HSC-}i$ & 1.2 & 0.59 & 26.5/25.7\phm{adf} & 26.2/25.6\phm{adf} & 1.0/1.8\phm{adf} \\
  & $\rm{HSC-}r$ & 0.6 & 0.59 & 26.7/25.8\phm{adf} & 26.3/25.7\phm{adf} & 1.0/1.8\phm{adf} \\
  & $\rm{HSC-}g$ & 0.6 & 0.72 & 27.0/26.2\phm{adf} & 26.5/26.0\phm{adf} & 1.0/1.8\phm{adf} \\\hline
GAMA15A & $\rm{NB964}$ & 85.3 & 1.03 & 25.2/24.6\phm{adf} & 24.6/24.3\phm{adf} & 1.2/1.8\phm{adf} \\
$14^{\rm{h}}22^{\rm{m}}00^{\rm{s}}$ & $\rm{HSC-}y$ & 1.2 & 0.80 & 25.0/24.4\tablenotemark{\rm{\footnotesize{psf}}} & 24.4/24.1\tablenotemark{\rm{\footnotesize{psf}}} & 1.0/1.8\tablenotemark{\rm{\footnotesize{psf}}} \\
$0^{\circ}00^{\prime}00^{\prime\prime}$ & $\rm{HSC-}griz$ & 3.6 & 0.61 & 27.0/26.0\phm{adf} & 26.6/25.9\phm{adf} & 1.0/1.8\phm{adf} \\
(J2000) & $\rm{HSC-}z$ & 1.2 & 0.71 & 25.8/24.9\phm{adf} & 25.3/24.8\phm{adf} & 1.0/1.8\phm{adf} \\
  & $\rm{HSC-}i$ & 1.2 & 0.58 & 26.5/25.7\phm{adf} & 26.2/25.6\phm{adf} & 1.0/1.8\phm{adf} \\
  & $\rm{HSC-}r$ & 0.6 & 0.54 & 26.6/25.7\phm{adf} & 26.3/25.7\phm{adf} & 1.0/1.8\phm{adf} \\
  & $\rm{HSC-}g$ & 0.6 & 0.72 & 27.0/26.2\phm{adf} & 26.5/26.0\phm{adf} & 1.0/1.8\phm{adf}
\enddata  
\tablenotetext{\rm{\footnotesize{psf }}}{ \:\,The $y-$band D $=1.8"$ $5\sigma$ depth measurements are computed for the HSC image that is PSF-matched to the NB DECam image. All other HSC depths are computed at native resolution.}
 \vspace{-0.7cm}
\label{btable}
\end{deluxetable*}

We observed LAGER fields WIDE12 and GAMA15A with the Blanco 4-m
telescope at Cerro Tololo Interamerican Observatory (CTIO) using the
Dark Energy Camera (DECam) instrument and the
narrow-band NB964 filter. NB964 has a central wavelength of 9642 \AA\ and
a narrow FWHM of 92 \AA\ that was custom-made to avoid bright skylines and
atmospheric absorption \citep{zheng19}. Given two
  surveys with the same limiting NB magnitude, a narrower FWHM has
  less bandpass dilution and is able to detect fainter emission line
  sources. A pure emission line source will be $\sim0.8$
  magnitudes brighter in our survey compared to a NB survey with twice our
  FWHM (e.g., \citealt{ota17}), and a Ly$\alpha$ emitter with a rest-frame EW$=10$ \AA\ will
be $\sim0.3$ magnitudes brighter in our survey. The DECam instrument is an
optical imager that has 62 CCDs covering a 2.2-degree diameter field
of view with a 0.264 arcsec pixel scale. Data were obtained over 20
nights and 6 observing semesters (NOAO PID: 2017A-0366; 2017A-0920;
2018A-0371; 2018B-0327; 2018B-0907; 2019A-0912). In total, we observed
WIDE12 for 100.6 ks and GAMA15A for 85.3 ks and obtained a $5\sigma$
point source survey depth of $\sim25$ mag with $1$ arcsec seeing
for both fields.  We used a $4'\times4'$ five-pointing
  `X' shaped dither pattern which we repeated, offsetting the entire pattern randomly by up to $30''$ for each new set of five exposures.

We supplement our NB964 data with overlapping Hyper Suprime-Cam Subaru
Strategic Program (HSC-SSP) $grizy$ DR2 broadband images \citep{aihara18}.
HSC is an optical imager on the 8.2m Subaru Telescope that has 104
CCDs covering a 1.5-degree diameter field of view with a 0.168 arcsec
pixel scale. The HSC-SSP is a three-layered multi-band survey that
consists of a Wide, Deep, and Ultradeep layer. Our LAGER WIDE12 and
GAMA15A fields are fully contained within the HSC-SSP Wide layer which
has 5 broadband images with $0.6$ arcsec seeing and a $5\sigma$
point source depth of $\sim26$ mag. We obtained HSC Pipeline \citep{bosch18}
calibrated broadband mosaic images via their public database\footnote{\url{https://hsc-release.mtk.nao.ac.jp}}.
We use the HSC $y$-band to select $y-$NB excess objects, and we
use the HSC $griz$-bands as veto bands to help reject low-redshift
interlopers.

\section{DECam and HSC Mosaic Images}

\subsection{DECam Image Stacking}

\noindent 

We collected all the WIDE12 and GAMA15A NB964 InstCal images associated
data quality maps (DQMs) observed before August 2019 from the NOAO
Science Archive\footnote{\url{http://archive1.dm.noao.edu/}}. NOAO
InstCal images are processed through the Community Pipeline (CP; \citealt{valdes14})
to remove instrumental effects and to perform an initial photometric
and astrometric calibration. For our specialized analyses, we require
astrometric and photometric calibrations beyond those produced by
the CP, and for this reason, we recalibrate using a procedure as described
in \citet{wold19}. Here we outline this astro-photometric calibration
and image stacking method with emphasis on alterations made to accommodate
our NB964 data.

\begin{figure*}[t]
\includegraphics[bb=84.15bp 49.5146bp 742.5bp 539.709bp,clip,angle=180,width=8.5cm]{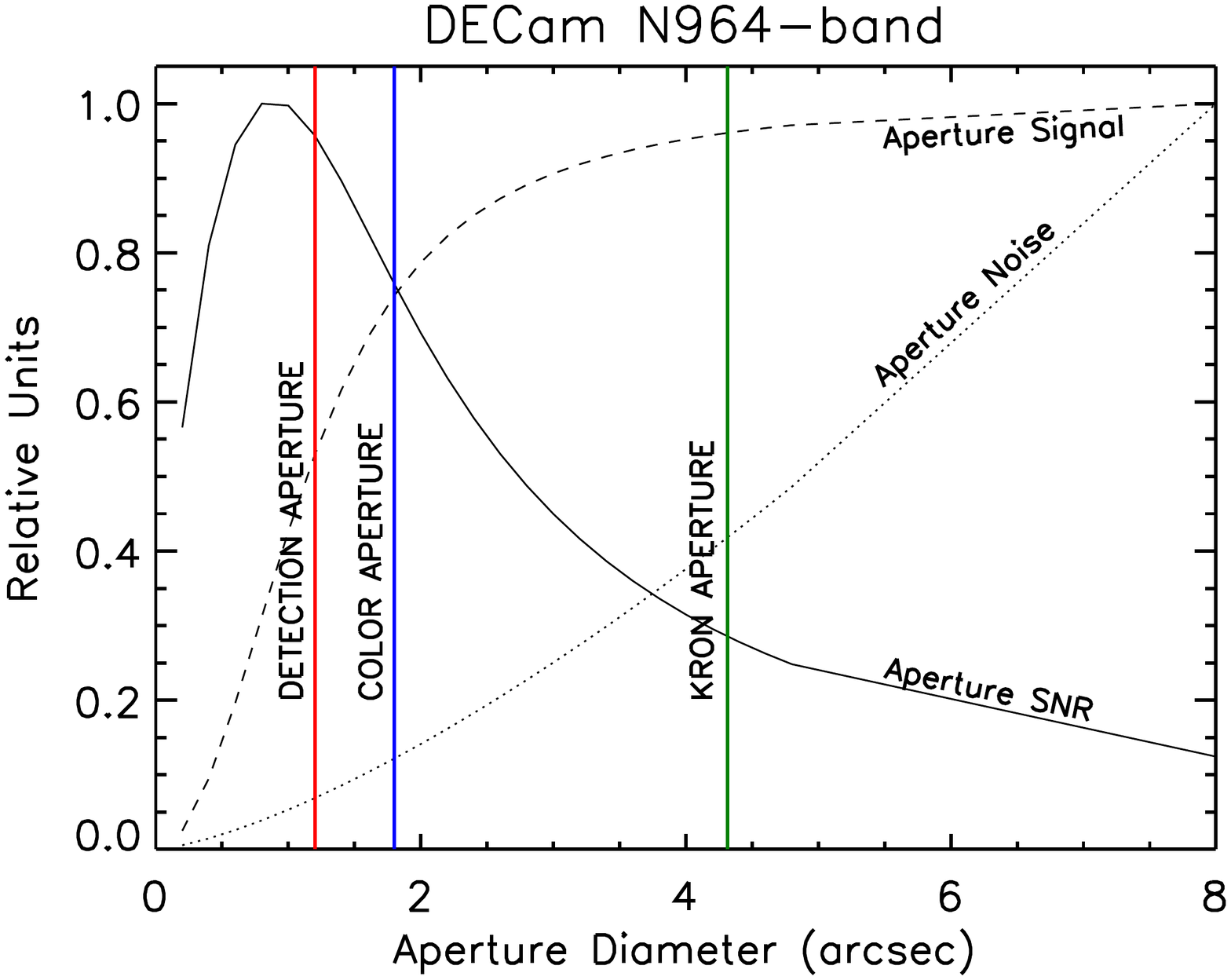}\includegraphics[bb=84.15bp 49.5146bp 742.5bp 539.709bp,clip,angle=180,width=8.5cm]{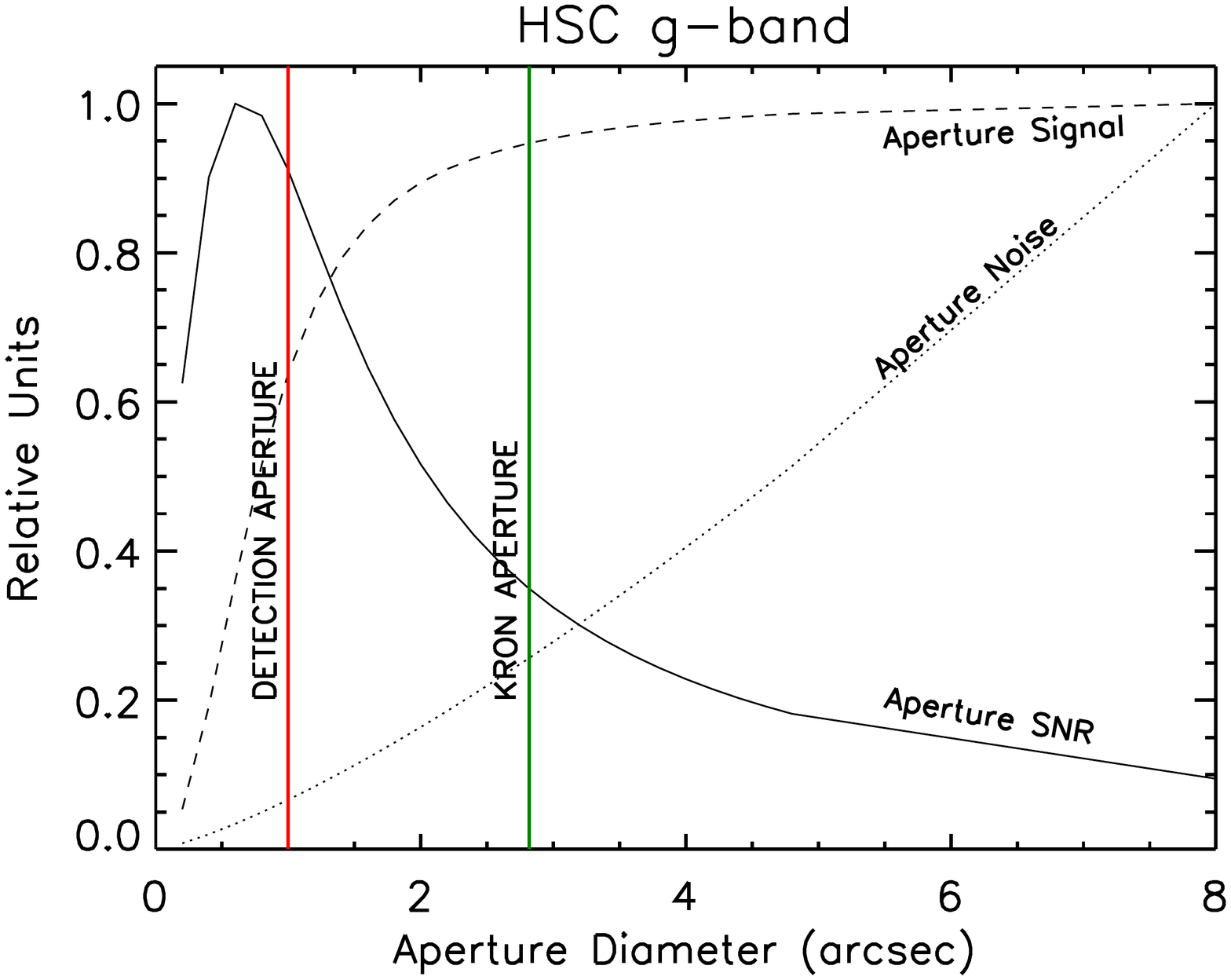}

\caption{Demonstration of our method for determining the optimal PSF aperture
for field WIDE12. The dotted, dashed, and solid curves show how the
noise, signal, and SNR vary with aperture size. For
  display purposes, all curves are normalized
to a maximum value of unity. The red vertical line shows our adopted
optimal extraction aperture, labeled as the `detection aperture'.
These apertures are slightly larger than the value at the peak of
the SNR curve to help mitigate systematic errors such as errors introduced
by non-perfect image co-alignment. The blue vertical line shows our
adopted aperture to measure $y-$NB964 colors, labeled as the `color
aperture'. As shown in Figure \ref{optimal_color}, we find that $1.8''$
apertures  
outperform 
Kron apertures when measuring $y-$NB964
colors. For comparison, the green vertical line shows the median SE
Kron aperture for point sources. The behavior for other filters and
for our other field, GAMA15A, are similar to the results shown here.
For both fields, we use a $1.0''$ detection aperture for the HSC
$griz$ bandpasses, and we use a $1.2''$ detection aperture for the
DECam NB964 bandpass. }

\label{optimal_ap}
\end{figure*}
Consistent with the HSC-SSP \citep{aihara18}, we use the Pan-STARRS1
(PS1) Data Release 2 catalog \citep{chambers16} as our astro-photometric
reference. PS1 is a $grizy$-band 30,000 deg$^{2}$ survey that has
millimag photometric and milliarcsec astrometric calibration. The
PS1 $grizy$-band 5$\sigma$ point source depths are 23.3, 23.2, 23.1,
22.3, and 21.4, respectively. For each DECam exposure, we identify
SNR $>10$ stars with no neighboring objects within $r=8''$ and determine
the median x- and y-offset needed to align our images to PS1. For
each exposure, we measure the seeing, the background rms, and the
relative flux scaling factors required to adjust all images to a designated
NB964 reference image.

With these measurements in hand, we produce image mosaics using SWarp
\citep{bertin02} employing point-source-optimized weighting \citep{gawiser06}.
When stacking, we assign zero weight to pixels flagged by the NOAO
DQMs which identify detector defects and image artifacts such as bleed
trails, saturation, and cosmic rays. Despite this practice, we find
that in some cases cosmic rays and satellite trails are not fully
masked in the resulting stacked image. Consequently, we employed our
own artifact removal procedure which is based on the algorithm presented
by \citet{gruen14}. To implement this procedure we produce a PSF-matched
median stack and flag pixels from individual PSF-matched science exposures
as bad if they significantly differ from the median stack. Defining
$f_{i}$ as the pixel flux of exposure $i$ and $\mu$ as the median
pixel flux of all exposures, we flagged pixels that met the following
criteria:

\begin{equation}
\left|f_{i}-\mu\right|>n\sigma_{i}+A\left|\mu\right|
\end{equation}

\noindent where $\sigma_{i}$ is the pixel noise and $n$ and $A$
are empirically determined clipping parameters that set the statistical
and PSF related leniency of the procedure. For our images which were
already flagged for artifacts with the NOAO pipeline, we used $n=5$
and $A=0.7$ to flag the remaining artifacts. Our procedure results
in revised DQMs that contain both the NOAO DQM flags
(masking $\sim3\%$ of image pixels) and our own artifact
flags (masking $\sim0.1\%$ of image pixels). We use these revised DQMs to flag bad pixels in non-PSF-matched
exposures when producing our final point-source-optimized image mosaics.

We photometrically calibrate our final NB mosaics using the PS1 $z$-
and $y$-bands which bookend our NB964 bandpass. As described in \citet{wold19},
we assume a linear color relation between PS1 and NB964 magnitudes
for point sources, such that:
\begin{equation}
{\rm {z}_{{\rm {PS1}}}={\rm {NB964}+\alpha(z_{{\rm {PS1}}}-y_{{\rm {PS1}}})+{\rm {ZPT}}}}
\end{equation}

\noindent We determine the color slope ($\alpha$) and the required
zero-point offset (ZPT) needed to adjust our NB magnitudes to AB by
solving for the best-fit line. We report the characteristics of our
NB mosaics and the accompanying broad band (BB) mosaics in Table \ref{btable}.

\subsection{Source Extraction}

We use SExtractor (SE; \citealt{bertin96}) in dual-image mode to
produce narrow-band selected catalogs. To run SE in double-image
mode, the NB detection image and the BB measurement image must have
the same dimensions. However, the HSC images have a smaller pixel
scale and a different pixel grid than our NB DECam mosaics. Given
this mismatch, we investigate two methods to produce the desired NB-selected
catalogs. 1) Following \citet{hu19}, we resample HSC's BB images ($grizy$-bands) to match
DECam's pixel scale and then generate the desired NB-selected catalog
from these co-registered images. 2) We create a synthetic NB detection
image with the same dimensions and pixel scale as the BB mosaics. Sources within
the synthetic NB detection image have Gaussian profiles
(FWHM $=2$ HSC pixels or $0.336''$) with coordinates determined
from the NB catalog. The main requirement for these
  inserted Gaussian sources is that SE can reliably detect and locate
  them, and this procedure is merely a stratagem to have SE perform
  aperture photometry on all the BB images at the
  coordinates of all NB objects. With this second method, we produce NB-selected catalogs
for the HSC $grizy$-band data without having to resample and thus
degrade the HSC data.

One potential drawback of this
 second method is the inability to use SE Kron flux measurements. Kron apertures are adaptive ellipses designed to
capture $\sim95\%$ of the total flux for both extended and unresolved
sources \citep{kron80,bertin96}. For LAGER fields COSMOS and CDFS, we used these apertures
to accurately measure $y-$NB colors for both extended and point-like sources.  In dual-image mode, Kron
apertures are based on the detection image morphology and by making a synthetic
 detection image we lose the NB morphology constraints. In Section \ref{select}, we compare
Kron colors measured from the resampled $y$-band image to aperture
colors measured from a $y$-band image PSF-matched to our NB mosaic.

To help identify foreground emitters, we also produced a point-source-optimized stack for
all bandpasses blue-ward of our NB filter, thus producing a combined
$griz$ master veto band.

As discussed in Section \ref{select}, we ultimately use the
  following NB-selected catalogs in the final LAE selection: 1) the
  non-resampled $g,r,i,z,griz$ veto band catalogs for elimination of foreground
  objects and 2) the PSF-matched $y$-band catalog for NB-excess measurements.

For all catalogs, we exclude objects that fall within our bright star
mask. The regions around bright stars have much higher
  background levels effectively masking out regions of the sky and
causing spurious detections due to diffraction spikes and saturation effects.
We mitigate these issues by masking out all 2MASS point sources \citep{cutri03}
with $J-$band magnitudes brighter than 13. Following the procedure
outlined in \citet{keenan10} and \citet{wold19}, we determined a
magnitude dependent circular star mask with the arc-second Radius
defined by:

\begin{equation}\label{bristar}
{\rm {Radius}=}326.3-41.2J+1.4J^{2}
\end{equation}

\noindent where the three coefficients are empirically determined
parameters set to remove spurious detections around bright stars.
With the star mask applied, we compute our survey area as $3.24$
deg$^{2}$ in WIDE12 and $2.91$ deg$^{2}$ in GAMA15A.

\section{LAE Candidates \label{select}}

\begin{figure}[t]
\includegraphics[bb=74.25bp 74.2718bp 717.75bp 460.485bp,clip,angle=180,width=8.5cm]{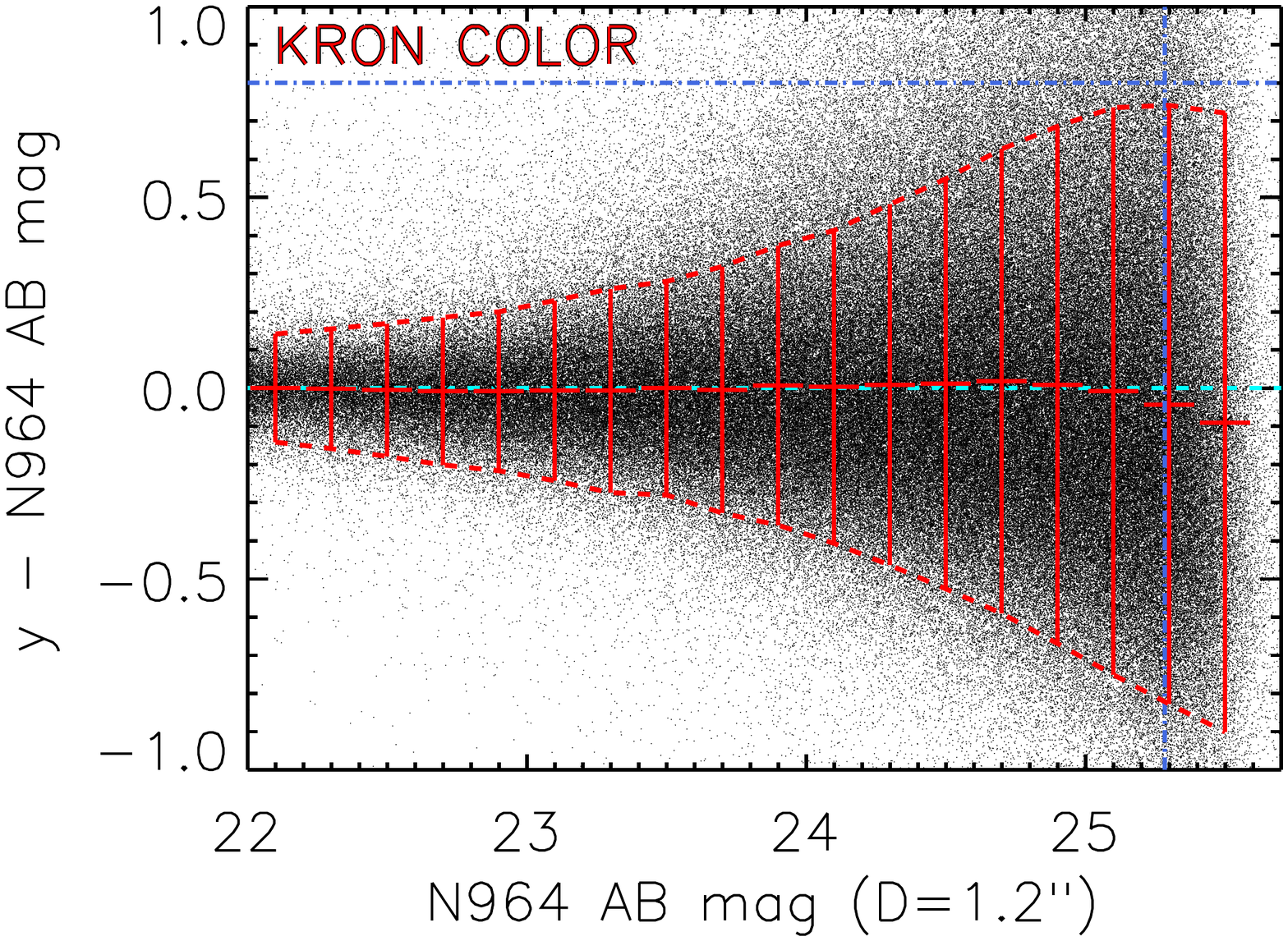}

\includegraphics[bb=74.25bp 79.2233bp 717.75bp 544.66bp,clip,angle=180,width=8.5cm]{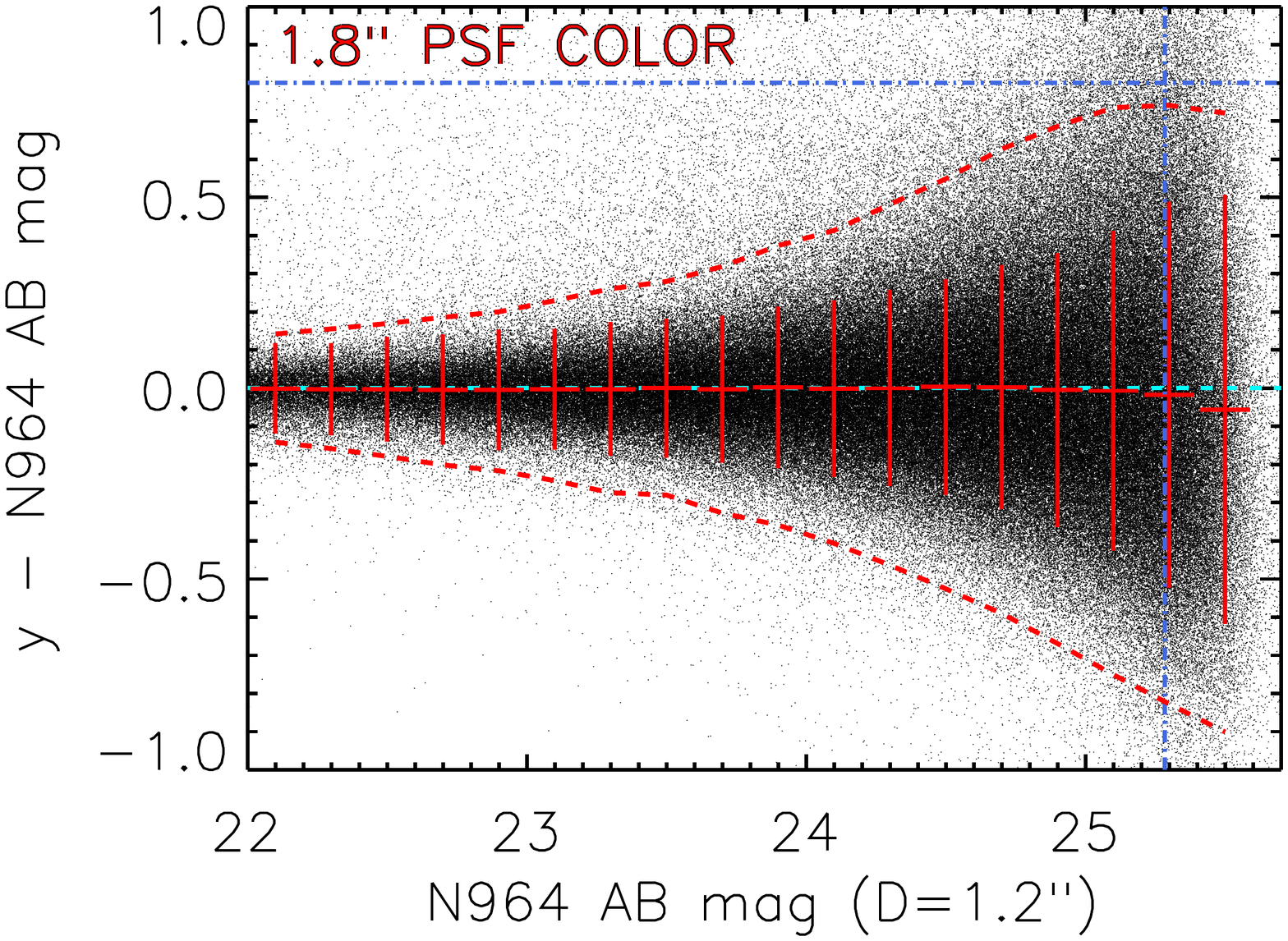}

\caption{(\textit{Top}) $y-$NB color determined with Kron apertures as a function
of NB D$=1.2''$ aperture magnitude for sources in WIDE12. The red
data-points show the binned median color and their error bars show
the standard deviation of the binned color measurements. The horizontal
blue dashed line shows the $y-$NB $>0.8$ mag color selection used
to isolate LAEs. The vertical blue dashed line shows the $5\sigma$
NB detection limit. (\textit{Bottom}) The same as the top, but colors
are measured with PSF-matched D$=1.8''$ apertures. The systematic
errors measured by the locations of the binned median colors are comparable
to the Kron results. However, the random error measured by the binned
color standard deviations are significantly reduced. For this reason,
we adopt PSF-matched $1.8''$ colors for our LAE selection in both
WIDE12 and GAMA15A fields (see Figure \ref{lae_select}). In both
the top and bottom figures, we show the envelope of the Kron color
data points (dashed red curves) for easy comparison between the two
results.}

\label{optimal_color}
\end{figure}
\begin{figure*}[t]
\includegraphics[bb=79.2bp 74.2718bp 697.95bp 544.66bp,clip,angle=180,width=8.5cm,origin=c]{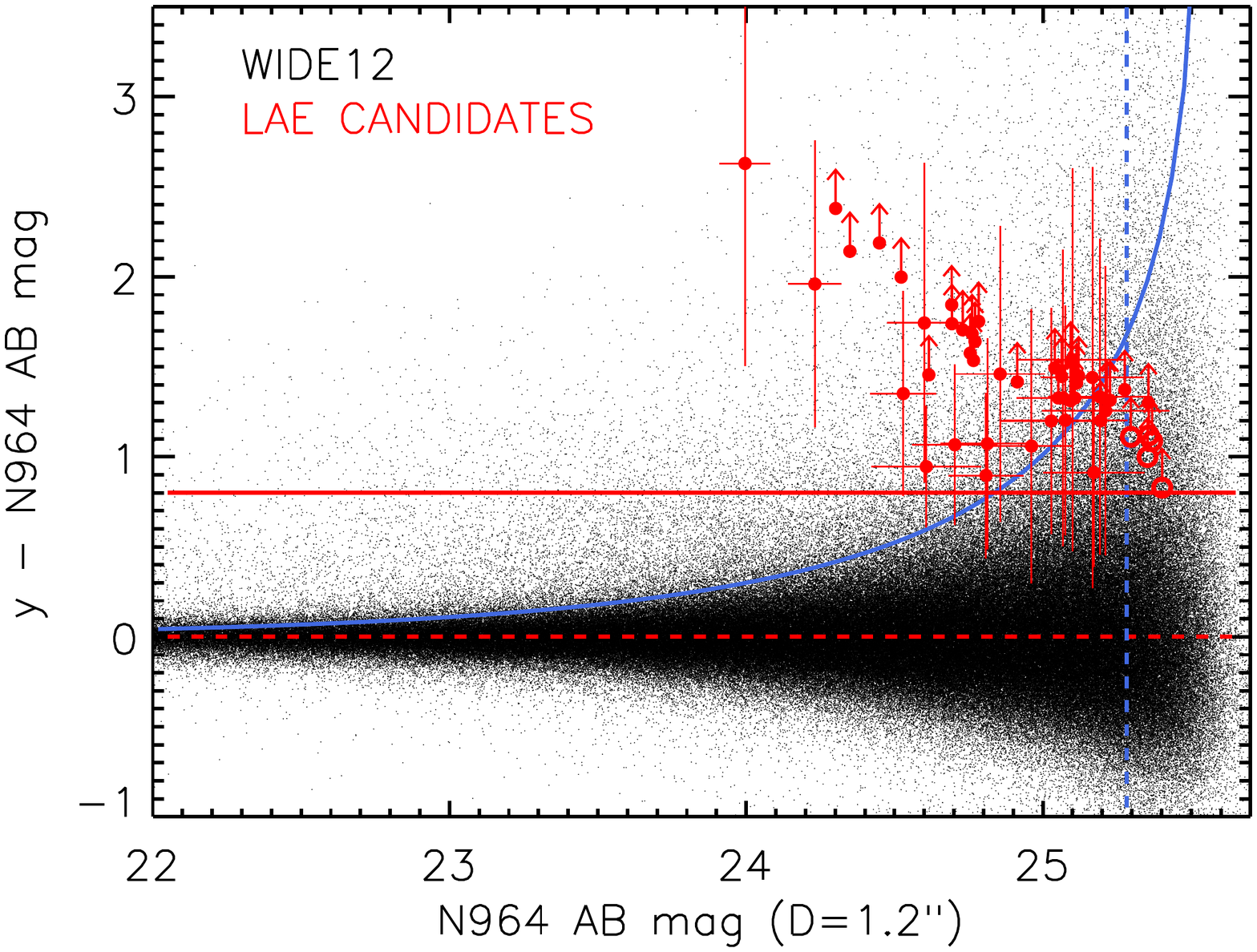}
\includegraphics[bb=79.2bp 74.2718bp 697.95bp 544.66bp,clip,angle=180,width=8.5cm,origin=c]{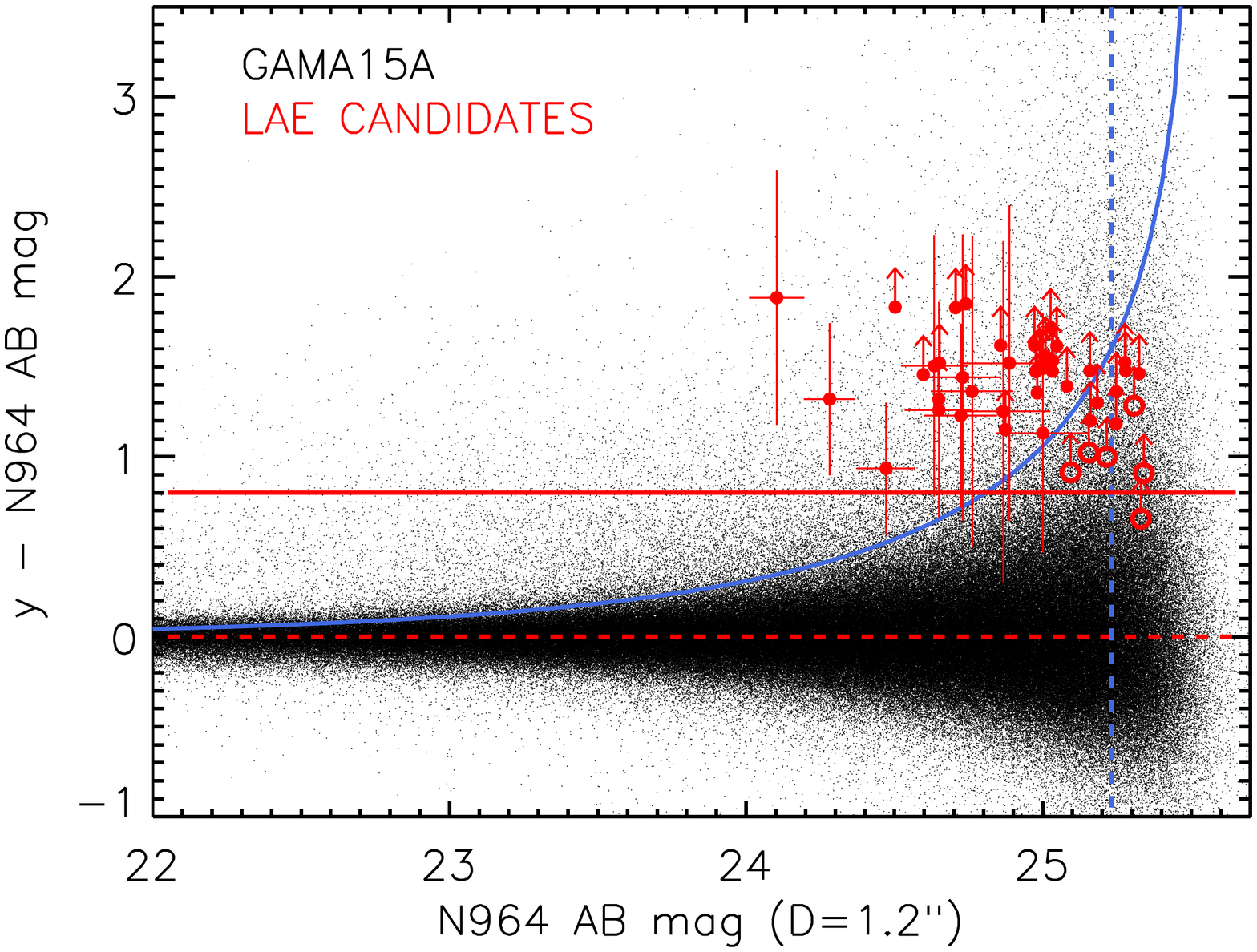}

\caption{To illustrate our LAE candidate selection, we show the NB color excess
as a function of NB D$=1.2''$ aperture magnitude for both WIDE12
and GAMA15A fields. The vertical dashed line indicates the median
$5\sigma$ depth of the NB image. The horizontal red line indicates
the 0.8 mag $y-$NB color cut. The blue solid curve indicates the
median $2\sum$ NB excess flux significance. LAE candidates (red data
points) satisfy all selection cuts as described in Section \ref{select}.
$y-$NB color lower limits (red arrows) are shown for candidates that
are not detected above $1\sigma$ in the $y$-band. Individual LAE candidates can fall below the
median selection cuts in the panels above, primarily in cases where
the local image depth is better than the median image depth.
LAE candidates that have completeness corrections less than
$5\%$ are excluded from our luminosity function computation and are
displayed above with open red circles (for details see Section \ref{comp})}.

\label{lae_select}
\end{figure*}
We wish to isolate a relatively small sample of $z=6.9$ LAEs from
our NB-selected catalogs which contain ${\rm N} \sim\hbox{530,000}$ objects per
field. The first step toward this goal is to identify the optimal
aperture size to measure the flux of our NB-selected objects. Of particular
interest are relatively compact sources, such as the targeted $z=6.9$
LAEs and one of our most challenging contaminants, high-EW foreground
emitters. To this end, we investigate which aperture size maximizes
the SNR for point sources. We estimate the signal as a function of
aperture size by measuring the median aperture flux relative to the
total flux in $0.2''$ intervals for our isolated star sample. In
other words, we compute the median curve of growth for point sources,
where $8''$ diameter apertures are used to measure the total flux. For
noise measurements, we randomly place ${\rm N}=\hbox{10,000}$ sky apertures within
our science images and compute the median absolute deviation of these
measured aperture fluxes (for a similar procedure see \citealt{gawiser06,wold19}).

In Figure \ref{optimal_ap}, we show how these quantities and their
ratio, the SNR for point sources, vary with aperture size. For both
fields, we adopt $1.2''$ diameter apertures to measure DECam fluxes
and $1.0''$ diameter apertures to measure HSC fluxes. We refer to
these apertures as our detection apertures. We chose apertures that
are slightly larger than the value at the peak of the SNR curve to
help mitigate systematic errors such as errors introduced by non-perfect
image co-alignment.

The second step toward isolating a sample of $z=6.9$ LAEs is to accurately
determine $y-$NB colors for both extended and unresolved sources.
\citet{hu19} showed that a color cut of $y-$NB $>0.8$ mag cleanly
isolates strong $\lambda_{{\rm {OBS}}}=9642$ \AA\ emitters (EW $\gtrsim10$
\AA\ for LAEs) from the more general continuum population. Ideally,
our adopted method for measuring color should maximize the $y-$NB
SNR while minimizing systematic errors that can be introduced by image
co-registration, differential seeing, and differential image depth.
To maximize the color SNR we would like to use apertures with sizes
similar to our detection apertures, and \cite{gawiser06} develop 
a method to use optimal detection apertures to measure colors for
both resolved and unresolved sources. This method derives aperture
corrections by estimating each object's intrinsic size from the detection
image. However, the relatively poor seeing of our NB detection image
compared to the HSC $y$-band seeing -- which in some cases results
in unresolved NB objects with resolved BB counterparts -- means that
we are unable to accurately measure intrinsic sizes and aperture corrections
using this method.

In Figure \ref{optimal_color}, we show the two methods that we considered
to determine $y-$NB colors: Kron aperture colors and PSF-matched
aperture colors. The main disadvantage of using Kron
apertures is their large size. For point sources within our NB mosaics,
the use of Kron apertures degrades the aperture SNR by a factor of
three relative to our detection aperture (see Figure \ref{optimal_ap}).

Our PSF-matched colors use smaller, more optimal apertures, at the expense
of having to smooth the measurement image to the detection image's
PSF. In Figure \ref{optimal_color}, we show that our PSF-matched
aperture size of $1.8''$ has systematic errors comparable to Kron
apertures but significantly reduces the $y-$NB color scatter. Given
the reduced color scatter shown in Figure \ref{optimal_color}, we
chose to use $1.8''$ PSF-matched colors over Kron aperture colors
in our LAE selection.

\begin{figure*}
\begin{centering}
\includegraphics[bb=100bp 10bp 895bp 755bp,clip,width=8.5cm]{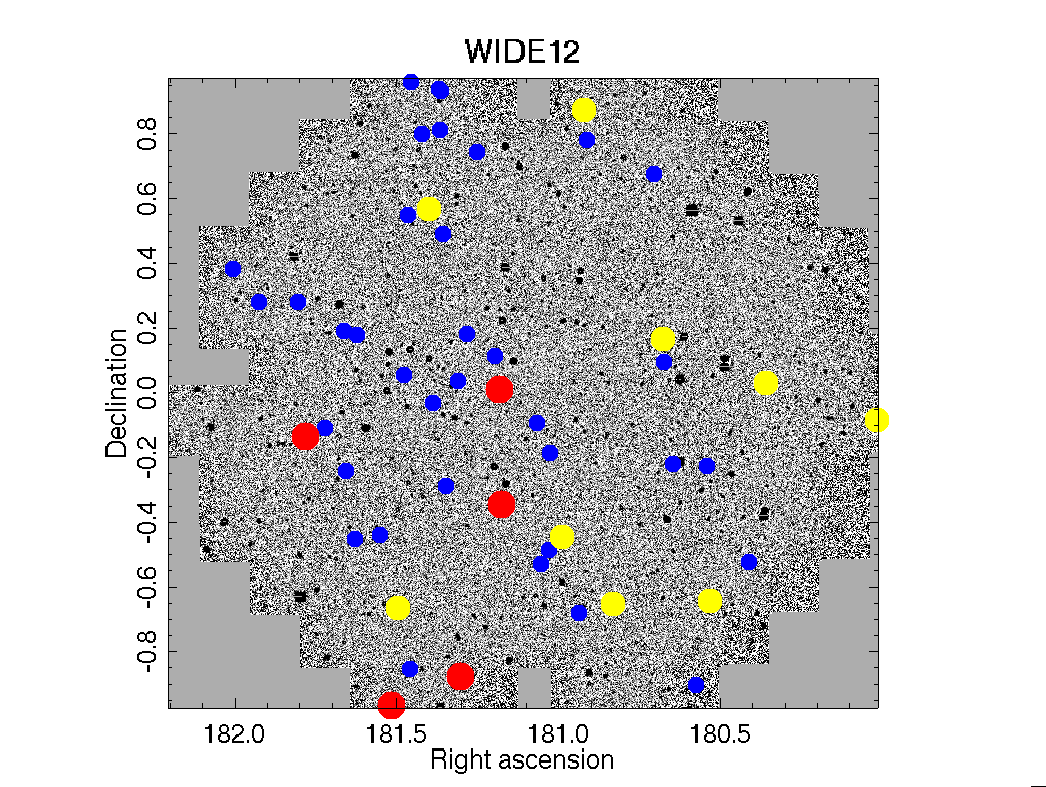}\includegraphics[bb=100bp 10bp 895bp 755bp,clip,width=8.5cm]{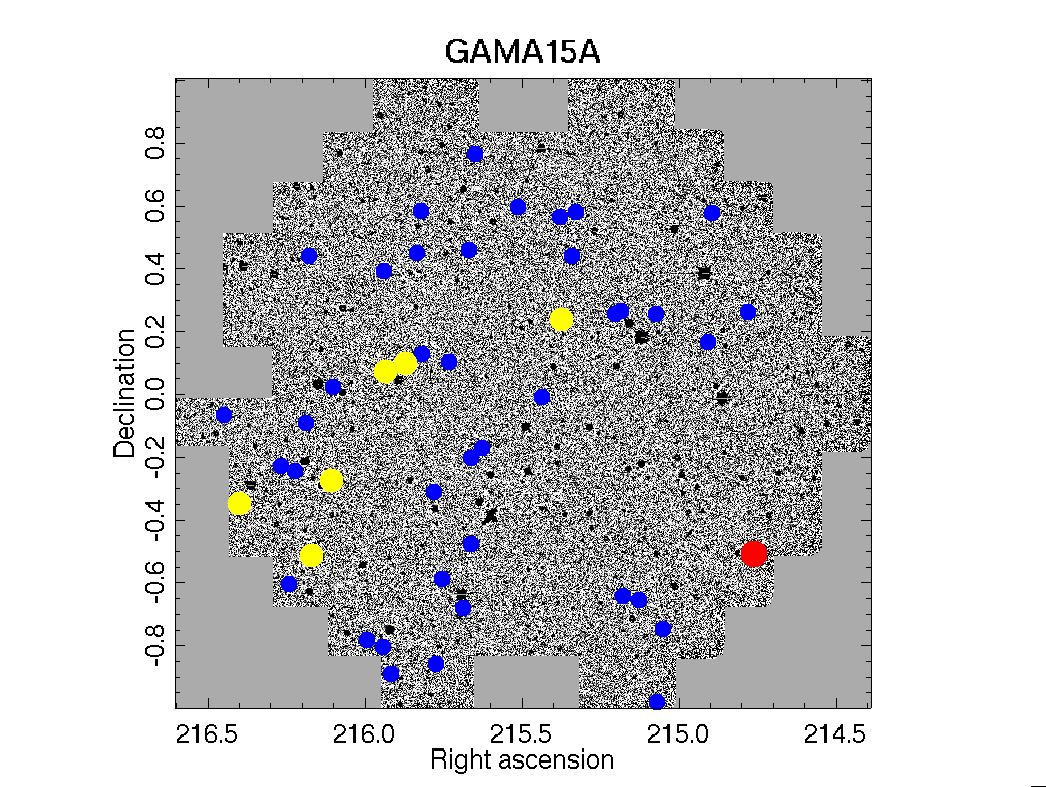}
\par\end{centering}
\caption{Spatial distribution of our $z=6.9$ Ly$\alpha$ emitters. Red, yellow,
and blue filled circles indicate LAEs with $L_{{\rm {Ly}\alpha}}>10^{43.3}$,
$10^{43.1-43.3}$, and $<10^{43.1}$erg s$^{-1}$. Unlike the LAGER
COSMOS field, we do not find bright LAEs preferentially in LAE over-densities.
This could reflect field-to-field variation, or our fields might not
be deep enough to detect the faint LAEs within the over-densities.
Our minimum Ly$\alpha$ luminosity is $\sim10^{42.8}$, while the
minimum luminosity for the CDFS and COSMOS fields is $\sim10^{42.6}$
erg s$^{-1}$. Both fields display $D\sim40'$ voids
that corresponds to $12.5$ pMpc at $z=6.9$.}

\label{fov}
\end{figure*}
For WIDE12 and GAMA15A, we select LAEs using $1.8''$ PSF-matched
apertures when relative measurements are needed and otherwise use
detection apertures via:

\begin{footnotesize}\begin{multline}\label{seleq}{\rm {SNR}_{1.2''}({\rm {NB964}})>5}\;\&\;{\rm     {SNR}_{1.0''}(\mathnormal{g,r,i,z,griz})<3}\;\&\;\\ \Sigma_{1.8''}>2\;\&\;\\ [(y_{1.8''}-{\rm {NB964}_{1.8''}>0.8}\;\&\;{\rm {SNR}_{1.8''}(\mathnormal{y})>3})\;{\rm {or}\;{SNR}_{1.8''}(\mathnormal{y})<3}] \end{multline}\end{footnotesize}

where

\begin{equation}
\Sigma_{1.8''}=\frac{f_{1.8''}({\rm {NB964}})-f_{1.8''}(y)}{\sqrt{\sigma_{1.8''}^{2}({\rm {NB964}})+\sigma_{1.8''}^{2}(y)}}
\label{sigeq}
\end{equation}

\noindent Our selection is similar to the previous LAGER studies \citep{hu19,zheng17}.
The only differences are the use of PSF-matched aperture colors rather
than Kron aperture colors, the explicit use of a combined $griz$
veto band, and the use of the $\sum$ parameter \citep{bunker95,sobral13,matthee15,coughlin18}
which measures the significance of the NB flux excess. In the previous
LAGER fields (COSMOS and CDFS) the available broadband images used
to determine the NB excess were 1-2 magnitudes deeper than the NB
images \citep{hu19}, and a $5\sigma$ NB detection with the (narrow $-$ broad) color 
of a line emitter was guaranteed a significant
NB flux excess. For our WIDE12 and GAMA15A fields, the broadband images
have about the same depth as the NB images (see Table \ref{btable}),
and the $\sum$ parameter is needed to guarantee a clean selection
of emitters at faint NB magnitudes. Negative $y$-band
fluxes can boost our $\sum$ parameter; however, we have verified that replacing negative y-band fluxes with zero does not change our final list of LAE candidates.

The overall filter set for the new fields is the same as for COSMOS.
Thus, based on analysis in fig.~2 of \cite{hu19}, 
the \lya\ equivalent width threshold for inclusion in our sample is 
EW$_{\rm{rest}} \ga 10$\AA\ (or  EW$_{\rm{obs}} \ga 80$ \AA),
with some dependence of the threshold on line wavelength (i.e.,
on precisely where the emission line falls within the filter transmission curve).
This is modified at the faintest NB964 magnitudes in the sample, where
the $\Sigma > 2$ criterion in equation~\ref{seleq} is
more stringent than the $y - {\rm {NB964}} > 0.8$ criterion, and the 
effective equivalent width threshold rises accordingly by a factor $\la 2$.
This can be seen in Figure~\ref{lae_select}, where
we illustrate the cuts used to isolate our LAE candidates.

The WIDE12 and GAMA15A veto bands are $\sim1$ magnitude shallower
than in the COSMOS and CDFS fields. To mitigate contamination from
faint foreground emitters, we combine all the available veto bandpasses
to construct a master $griz$ veto image and require our candidates
to be undetected at the $3\sigma$ limit (see Section \ref{pure}
for discussion of the purity of our sample).

We find $149$ and $131$ LAE candidates in our WIDE12 and GAMA15A
fields, respectively. Visual inspection was performed independently
by three of the authors. All LAE candidates were visually verified
by inspecting all available bandpasses ($g,r,i,z,griz,y,NB$) for
potential problems. The three visual classifications were found to
largely agree. Artifacts, such as diffraction spikes and cosmic rays,
contaminated our candidate list at the $25\%$ level, while sources
with weak counterparts in one of the veto bands but below our formal
$3\sigma$ cut contaminated our candidate list at the $41\%$ level.
We find a final clean sample of $N=50$ and $N=45$ LAE candidates
in the WIDE12 and GAMA15A fields, respectively. This more than doubles
the LAGER LAE sample size from $N=79$ \citep{hu19} to $N_{{\rm {4\text{-}Field}}}=174$.

We compute Ly$\alpha$ line fluxes from the aperture corrected NB
and BB $D=1.8"$ fluxes. For LAEs with undetected BB fluxes, we use
$1\sigma$ $y-$band measurements to estimate BB flux. We assume the
LAE candidates have a $\delta-$function Ly$\alpha$ line profile
at the center of the NB filter and a UV continuum slope of $-2$ that
is attenuated by the IGM via the \citet{inoue14} model. Given this
assumed spectral shape, the NB/BB filter responses, and the NB/BB fluxes
we solve for the normalization of the UV continuum and the Ly$\alpha$
flux \citep[for a similar procedure see][]{hu19}.

In Figure \ref{fov}, we show the spatial distribution of our final
LAE candidate sample for both fields. Within our two fields, we find
about the same number $N=6$ of bright log$(L_{{\rm {Ly}\alpha}})>43.3$
erg s$^{-1}$ LAE candidates (red filled circles) as found by Hu et al.\ in LAGER fields
COSMOS and CDFS where $N=7$. Furthermore, we notice large field-to-field
variation with both the WIDE12 and COSMOS fields displaying a higher
concentration of bright LAEs relative to the GAMA15A and CDFS fields.
Unlike what was found in the LAGER COSMOS field, we do not find bright
LAEs preferentially in LAE over-densities. This could reflect field-to-field
variation, or our fields might not be deep enough to detect the faint
LAEs within the over-densities. Our minimum Ly$\alpha$ luminosity
is $\sim10^{42.8}$, while the minimum luminosity for the CDFS and
COSMOS fields is $\sim10^{42.6}$ erg s$^{-1}$. 

\section{Reliability of the LAE Sample\label{pure}}

We estimate the foreground contamination in LAE samples from the
new  WIDE12 and GAMA15 fields using deeper images from the LAGER
COSMOS field.  
We ask two related questions.
(a) First, how many emission line galaxies in COSMOS have measured veto filter fluxes
lying between the $3\sigma$ limit for COSMOS and the brighter $3\sigma$ limit
for WIDE12 and GAMA15?  Such galaxies would be LAE candidates in our shallower
fields, but would be correctly ruled out as foreground emitters given deeper data.
(b) Second,  if we randomly perturb the measured fluxes in the COSMOS emission line catalog,
adding noise to simulate the depth of the shallower fields, how many other galaxies will scatter 
into the LAE selection region?

COSMOS has archival HSC veto bandpass
data typically deeper by $1-2$ magnitudes, along with published photo-$z$
measurements based on 30-band photometry \citep{laigle16}. By using
the existing COSMOS catalog of foreground emitters \citep{khostovan20}
and artificially degrading the depth to match the WIDE12 and GAMA15
depths, we can estimate the number of foreground emitters contained
within the WIDE12 and GAMA15 LAE candidate lists.

We begin with Khostovan et al.'s NB964-selected
  COSMOS catalog of ${\rm N}=\hbox{10,877}$ EW$_{\rm{OBS}}>52.2$ \AA\ H$\alpha$,
{[}O{\small{}III}{]}, and {[}O{\small{}II}{]} emitters which is photo-$z$
and color selected based on Laigle et al.'s COSMOS2015 catalog ($B,r,i,z,y$
with $5\sigma$ depths of $26.4, 25.9, 25.6, 25.3, 24.2$). Khostovan et
al.'s catalog selects emitters down to a $5\sigma$ NB depth of $25.45$.
This is $\sim0.8$ magnitudes deeper than our WIDE12 and GAMA15A
fields, where the total $5\sigma$ depth is $24.6$.  Furthermore, the NB
magnitude distribution for the selected foreground emitters peaks at $\sim24.5$, which is comparable to our WIDE12/GAMA15 NB depth limit.

We cross-match this foreground emitter catalog to
the HSC Subaru Strategic Program DR2 catalog ($g,r,i,z,y$ with $5\sigma$ depths of
$27.3, 26.9, 26.7, 26.3, 25.3$).  We perform this cross-match for two
reasons: 1) our LAE selection employs HSC bandpasses and 2) the HSC DR2 survey
is $\sim1$ magnitude deeper than Laigle et al.'s COSMOS2015 catalog.

We find the closest cataloged HSC DR2 match within a $1.5''$ search radius of the emitter's
coordinate. We accept the match if a $\left|r_{{\rm {Laigle}}}-r_{{\rm {HSC}}}\right|<1$
mag or $r_{{\rm Laigle}}>26$ mag, where the $5\sigma$ depth of $r_{{\rm {Laigle}}}$
is $\sim26$ mag. Our adopted condition for accepting
  a cross-match is designed to default to the Laigle et al.\
  catalog  -- the catalog used to identify foreground emitters --
  when there are significant conflicts between the two surveys. We find that $90\%$ of the foreground emitters have an accepted match,
and record their HSC magnitudes. For the remaining $10\%$, we record
their Laigle et al.\ magnitudes.  Visually inspecting the
  objects with no accepted HSC match, they appear to be dominated by
  HSC DR2 background subtraction issues around bright stars and deblending
  issues.  For these objects, we regard the Laigle et al.\ catalog to
  be more reliable, and our catalog cross-matching method ensures that
  the Laigle et al.\ flux values are used.

For the purposes of this foreground contamination estimate, we use our field's
total bandpass depths  (see Table \ref{btable}) and HSC DR2's CModel
total magnitudes which are designed to measure total fluxes for both
extended and point sources \citep{bosch18}. For the small percentage
of sources with only COSMOS2015 magnitudes, we use their cataloged
Kron magnitudes to measure total fluxes.

Using this final catalog of foreground emission line galaxy fluxes, we find
five galaxies that are securely identified as foreground emitters in COSMOS, but whose
measured veto band fluxes would fall below the $3\sigma$ threshold given the 
shallower broad-band imaging depths of either the WIDE12 or the GAMA15 field.  

To further explore the impact of photometric errors on our sample
contamination, we perform a simple simulation to see how many galaxies scatter into
our LAE selection region in color space.  For each object in the input (COSMOS) catalog, 
we perturb the photometry, adding an amount of random noise to reproduce the 
photometric uncertainties of WIDE12 and GAMA15.  
We independently perturb the test catalog 10,000 times, each time applying our LAE
  selection criteria (Equation \ref{seleq}). Averaging over all the
  runs, we find $5.1\pm1.5$ contaminants for WIDE12 and $5.6\pm1.6$
  contaminants for GAMA15A. These foreground emitters typically have NB-magnitudes ranging from
$23.5-24.1$ and high equivalent widths (EW$_{{\rm
    {OBS}}}$$>180$\AA).

We find $N=45$-$50$ LAEs per
  new LAGER field and this indicates a  $f_{{\rm {cont}}}\sim11\%$
  contamination rate. However, the COSMOS LAE selection is performed over a slightly smaller area
when compared to our current fields. Correcting for our slightly larger
area, 

our Monte Carlo procedure yields a $\pm 1\sigma$ range of
 $f_{\rm{cont}}=10-19\%$ contamination in our
WIDE12 and GAMA15A LAE sample.

\begin{figure}[t]
\includegraphics[bb=79.2bp 89.1262bp 688.05bp 544.66bp,clip,angle=180,width=8.5cm]{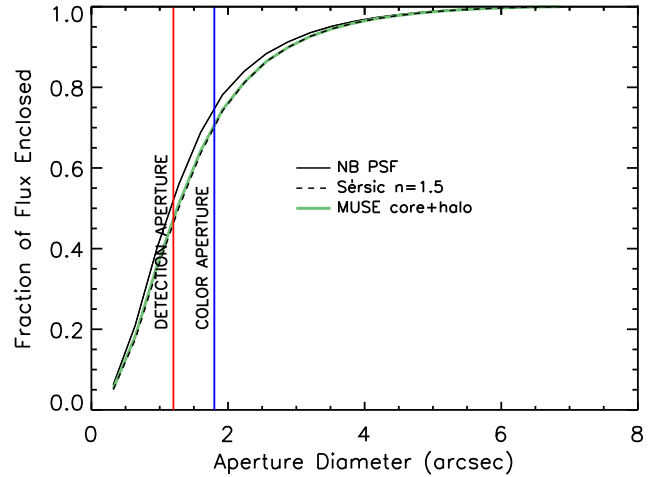}

\caption{Comparison between our simulated MUSE-inspired profile (green curve)
and the S\'ersic n=1.5 profile used by recent $z\sim7$ LAE surveys
(black dashed curve, \citealt{konno18,hu19}). Both are convolved
with our NB PSF (black solid curve). Given the similarity between
these spatial profiles, we simulate LAEs with a S\'ersic n=1.5 profile
to be consistent with past studies.}

\label{lae_psf}
\end{figure}

\begin{figure*}[t]
\begin{centering}
\includegraphics[bb=108.9bp 89.1262bp 693bp 539.709bp,clip,angle=180,width=8.7cm]{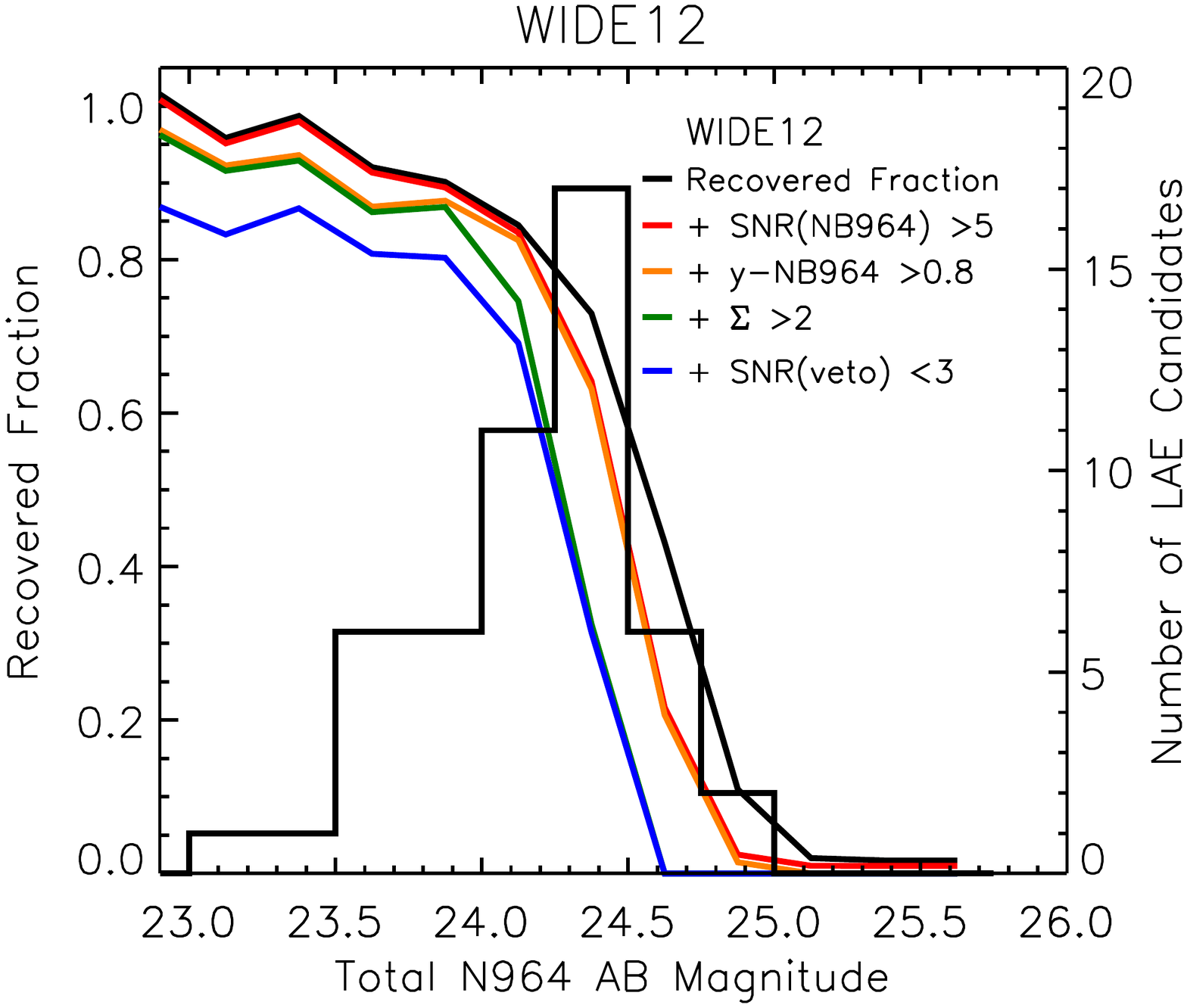}\includegraphics[bb=108.9bp 89.1262bp 702.9bp 539.709bp,clip,angle=180,width=8.7cm]{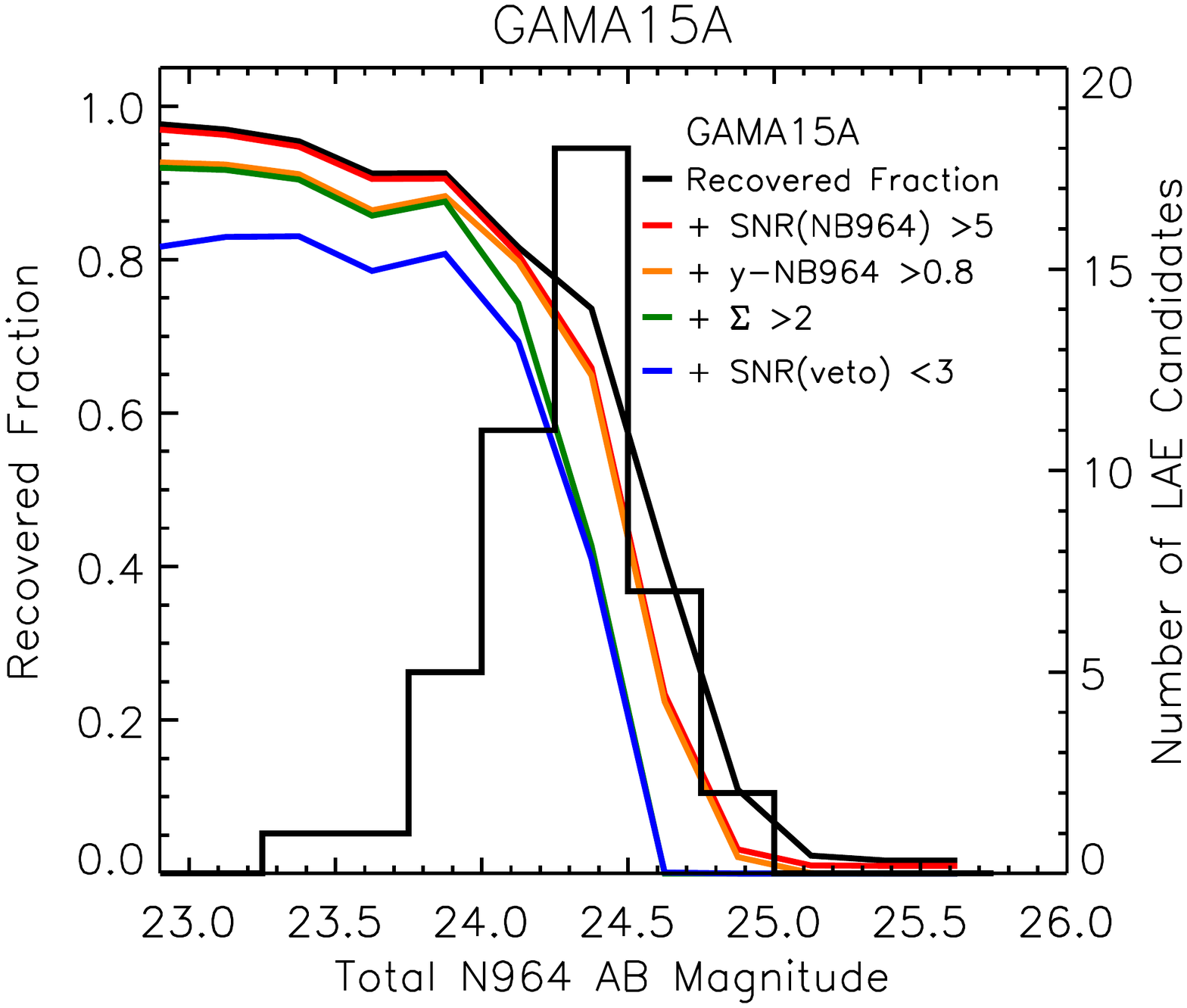}
\par\end{centering}
\caption{WIDE12 and GAMA15A LAE completeness curves as a function of total
NB magnitude. We show the effect of sequentially applying our LAE
selection cuts with the red, orange, green, and blue completeness
curves. The black histogram indicates the NB magnitude distribution
of our LAE candidates. By convention, simulated sources are not prevented from
falling within the isophotes of real sources. Thus, the measured flux
from the recovered source can be boosted. Additionally, inputted sources
have their NB and BB flux altered by the background noise which can
cause sources to scatter into adjacent magnitude bins. In practice
our total completeness (blue curve) never exceeds $\sim85\%$. This
is primarily caused by our veto band cuts which remove all simulated
LAEs that happen to fall within a $3\sigma$ isophote of a foreground
source.}

\label{1d_complete}
\end{figure*}

Our foreground contamination estimate
neglects any field-to-field variation by assuming the COSMOS field
is representative of our other LAGER fields. We emphasize that spectroscopic follow-up has been very
successful with an $80\%$ recovery rate found in a Keck/LRIS follow-up
of $21$ LAE candidates \citep{harish21}.  Based on our best
catalog-based estimate and based on our on-going follow-up, we expect some moderate level of contamination in our LAE sample which
is consistent with archival Ly$\alpha$ surveys (e.g., \citealt[][estimate $f_{\rm{cont}}=17-33\%$]{konno18}).

\section{Completeness of the LAE Sample \label{comp}}

We insert artificial LAEs into our NB and BB science images and then
extract, measure, and select Ly$\alpha$ candidates using our standard
procedure explained in Sections \ref{select}. We use the number of
recovered objects over the number of input objects per NB magnitude
bin to estimate our sample's completeness.

For completeness measurements, the MUSE LAE surveys at $z=3-6$ \citep{wisotzki16,leclercq17,herenz19}
have emphasized the need to account for the LAE's extended spatial
profiles. These studies found significant LAE diffuse halos which,
if not properly accounted for, can result in a factor of $\sim2$ under-estimates
of the faint-end of the LAE LF. Furthermore, Wisotzki et al.\ find
that higher-redshift LAEs tend to have smaller halo scale lengths
and higher-EW LAEs tend to have lower halo flux fractions. Given these
trends and our high-redshift, high-EW LAE sample, we made simulated
LAEs with 1) a halo exponential scale length of 1 kpc, 2) a core exponential
scale length of $0.1$ kpc, and 3) a halo flux fraction of $0.4$.
We expect the LAE population to display a distribution of spatial
profiles. However, observational constraints on this distribution
and potential dependences on physical quantities are lacking especially
at the highest redshifts. Given these uncertainties, we used a single
spatial profile in our completeness simulations.

In Figure \ref{lae_psf}, we show that there is little difference
between our simulated MUSE-inspired profile and a S\'ersic n=1.5
profile with a half-light radius of $0.9$ kpc used by previous high-redshift
LAE surveys \citep{konno18,hu19} once convolved with our NB PSF.
Given this result and for consistency with past studies, we simulate
LAEs with a S\'ersic n=1.5 profile for both their NB and $y-$band
counterparts. We also base our $1.8''$ color aperture
  corrections ($-0.38$ and $-0.32$ mag for WIDE12 and GAMA15A, respectively) on this profile
when computing Ly$\alpha$ fluxes for our LAE samples. This adopted S\'ersic profile is similar to UV continuum profile
measurements of LAE and LBG galaxies in the epoch of reionization
\citep{jiang13,allen17,shibuya19}.

For our completeness simulations, the image positions of our artificial
LAEs were randomly selected excluding regions flagged by our bright
star mask (see Equation \ref{bristar}). To gain more statistical leverage at fainter fluxes, the
simulated LAE NB fluxes, $f_{{\rm {\rm {NB}}}}$, were randomly sampled
from a power law distribution ($dN/df_{{\rm {NB}}}\propto f_{{\rm {NB}}}^{-2.5}$)
with a minimum flux threshold of 26 mag. The $y-$band fluxes were
assigned by randomly sampling an exponential EW distribution with
a rest-frame scale length of $100$\AA\ and a minimum rest-frame EW
threshold of $5$\AA. After inserting artificial sources, we performed
our standard LAE selection and measured 
the fraction of all simulated EW $>10$\AA\ objects
successfully recovered, as a function of NB flux.  This procedure
allows us to correct for any EW incompleteness that
  is introduced by our $2\sum$ and NB-excess cuts.

In Figure \ref{1d_complete}, we plot the recovered
fraction of artificial sources as a function of total NB magnitude.
We show the effect of sequentially applying our LAE selection cuts
with the red, orange, green, and blue completeness curves. Our NB
SNR cut removes faint LAEs not detected at $5\sigma$ significance
(red curve). Our NB-excess cut removes LAEs with EW $\lesssim10$
\AA\ (orange curve). Our $\sum$ parameter cut removes sources without
a flux excess detected at $2\sigma$ significance (green curve). This
cut preferentially removes low-EW LAE candidates at faint NB magnitudes. Finally,
our $3\sigma$ veto band cuts remove simulated LAEs that are randomly
positioned on the isophotes of foreground objects.

As discussed in detail by \cite{hu19} 
 (see their discussion of ``selection
incompleteness''), even for NB bright LAEs we do not recover $100\%$
of our inputted LAEs. 

Both fields have a maximum total completeness of $\sim83\%$
as shown by the blue curve in Figure \ref{1d_complete}. 
This is mainly due to our veto band
cuts that mask out any regions falling within the $3\sigma$ isophotes
of foreground sources. For both of our fields, we find that $\sim13\%$
of randomly placed $D=1.0''$ apertures have $griz$-band flux measurements
above $3\sigma$.  The remaining deficit ($\sim 4\%$) is due to incomplete recovery of LAE candidates 
in the $10{\rm \AA} \la EW \la 20 {\rm \AA}$ range.

In Figure \ref{1d_complete}, we also show the NB magnitude
distribution of our LAE candidates. 
The number of candidates rises with decreasing flux
from the brightest LAEs in the field (at total AB 
magnitude $\hbox{\rm NB964}\sim 23$) to a peak at  $\hbox{\rm NB964}\sim 24.4$,
indicating the rising luminosity function. Fainter than   $\hbox{\rm NB964}\sim 24.5$,
it falls off with the declining completeness near the survey limit.

When calculating luminosity functions, 
we wish to reach the faintest fluxes where our data can provide
useful measurements.  In practice we select our completeness
threshold such that the bin-averaged completeness in our faintest
luminosity function bins is $\sim 25\%$.  This is comparable to 
other recent $z\approx 7$ \lya\ luminosity function studies 
(e.g., \citealt{ota17} report a minimum binned completeness 
of $\sim22\%$ and $\sim35\%$ for their SDF and SXDS fields).
We have therefore chosen to use all LAEs with individual completeness $>5\%$, 
which yields bin-averaged completeness of 22\% in the faintest bin
of the WIDE12 luminosity function, and 30\% in GAMA15A.
This cut excludes 11 LAE candidates (5 in WIDE12, and 6 in GAMA15A), leaving
us with a final sample size of $N=84$ LAEs that are used in luminosity 
function calculations in section~\ref{sec:lfcalc}.  For all figures, tables, and analyses, we scale
our Poisson LF error bars by our incompleteness.

\begin{deluxetable*}{ccccccc} 
\tablecolumns{7} 
\tablewidth{0pc} 
\tablecaption{Best-fit $z=6.9$ Ly$\alpha$ Luminosity Function Parameters}
\label{lftable}
\tablehead{ 
\colhead{Field} \vspace{-0.2cm} & \colhead{log $L_{\rm{Ly}\alpha}$ Fitted Range} & \colhead{$\alpha$} & \colhead{$L^{*}$} & \colhead{$\phi^{*}$} & \colhead{$[\rho_{\rm{Ly}\alpha}]_{42.4}^{\infty}$}  & \colhead{Displayed in}\\
\colhead{ } & \colhead{(erg s$^{-1}$)}  & \colhead{} & \colhead{($10^{42}$ erg s$^{-1}$)} & \colhead{($10^{-4}$Mpc$^{-3}$)}& \colhead{($10^{39}$erg s$^{-1}$ Mpc$^{-3}$)} & \colhead{ }}
\startdata 
WIDE12 & 42.87-43.52 & -2.5(fixed) & 9.31$^{+7.79}_{-3.17}$ & 2.24$^{+5.98}_{-1.75}$ &  2.71$^{+0.81}_{-0.82}$ & Figure \ref{wide_gama}, \ref{indiv_lf} \\
 & 42.87-43.52 & -1.7(fixed) & 6.03$^{+2.60}_{-1.53}$ & 5.20$^{+8.13}_{-3.11}$ &  2.03$^{+0.55}_{-0.64}$ & Figure \ref{wide_gama} \\
\hline
GAMA15A & 42.86-43.37 & -2.5(fixed) & 6.25$^{+5.17}_{-2.16}$ & 6.44$^{+23.42}_{-5.34}$ &  3.21$^{+1.17}_{-0.99}$ & Figure \ref{wide_gama}, \ref{indiv_lf} \\
 & 42.86-43.37 & -1.7(fixed) & 4.42$^{+2.10}_{-1.13}$ & 11.11$^{+21.25}_{-7.43}$ &  2.43$^{+0.71}_{-0.85}$ & Figure \ref{wide_gama} \\
\hline
WIDE12+ & 42.86-43.52 & -2.5(fixed) & 7.05$^{+2.77}_{-1.70}$ & 4.52$^{+6.44}_{-2.79}$ &  2.98$^{+0.64}_{-0.75}$ & --\\
GAMA15A & 42.86-43.52 & -1.7(fixed) & 4.98$^{+1.42}_{-0.95}$ & 7.97$^{+7.88}_{-3.98}$ &  2.19$^{+0.41}_{-0.59}$ & -- \\
\hline
LAGER 4F & 42.69-43.54 & -2.5(fixed) & 8.95$^{+2.82}_{-1.79}$ & 2.19$^{+1.79}_{-1.09}$ &  2.44$^{+0.31}_{-0.36}$ & Figure \ref{combo_lf}, \ref{comp_lf} \\
 & 42.69-43.54 & -1.7(fixed) & 5.65$^{+1.08}_{-0.87}$ & 5.55$^{+3.36}_{-1.88}$ &  1.93$^{+0.28}_{-0.27}$ & Figure \ref{combo_lf} \\
 & 42.69-43.26 & -2.5(fixed) & 5.92$^{+1.94}_{-1.14}$ & 6.99$^{+7.47}_{-4.10}$ &  3.07$^{+0.41}_{-0.52}$ & Figure \ref{combo_lf} \\
 & 42.69-43.26 & -1.7(fixed) & 4.01$^{+0.77}_{-0.61}$ & 14.03$^{+9.41}_{-5.81}$ &  2.52$^{+0.38}_{-0.37}$ & Figure \ref{combo_lf} \\
\enddata  
\tablecomments{}
 \vspace{-0.9cm}
\end{deluxetable*}

\noindent 
\begin{figure}
\includegraphics[bb=99bp 99.0291bp 712.8bp 544.66bp,clip,angle=180,width=8.5cm]{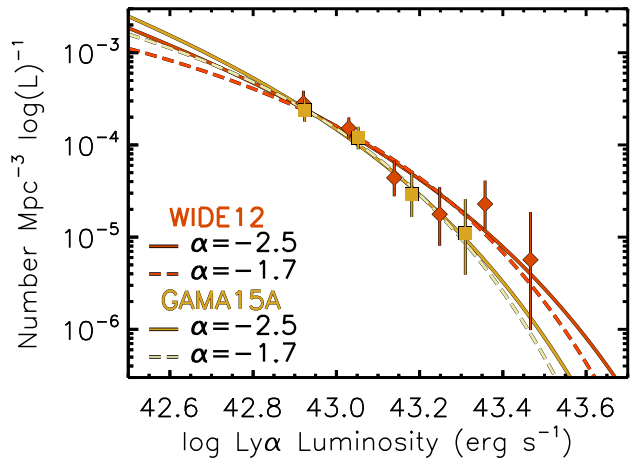}

\caption{The LAGER WIDE12 (orange diamonds) and GAMA15A (yellow squares) Ly$\alpha$ LFs and best-fit Schechter
functions. Field WIDE12 displays a more prominent bright-end
($L_{{\rm {Ly}\alpha}}>10^{43.3}$ erg s$^{-1}$) tail compared to
GAMA15A. An analogous result was also found between our two previous
LAGER fields, COSMOS and CDFS, where COSMOS was found to have an over-dense
bright-end.} 

\label{wide_gama}
\end{figure}

\section{Ly$\alpha$ Luminosity Functions at $z=6.9$}
\label{sec:lfcalc}

\nopagebreak[4]

\subsection{The WIDE12 and GAMA15A Ly$\alpha$ LFs}

\nopagebreak[4]

To explore field-to-field variations, we compute the
  WIDE12 and GAMA15A $z=6.94$ Ly$\alpha$ LF using the $1/V_{{\rm {max}}}$
technique \citep{felten76}. In Figure \ref{wide_gama}, we show our
EW $>10$\AA\ Ly$\alpha$ luminosity function for both WIDE12 and
GAMA15A. We find that WIDE12 displays a more prominent bright-end
tail compared to GAMA15A. This is similar to what was found in our
previous LAGER fields, where COSMOS was found to have a pronounced
bright-end bump compared to CDFS. 
Applying a Kolmogorov-Smirnov test to the bright end LFs for the 
four fields ($\log(L) > 43.04$ erg/sec) confirms the COSMOS bright-end
excess at confidence levels from $98\%$ (compared to WIDE12) to 
$99.8\%$ confidence (compared to GAMA15), while the bright end of 
the WIDE12 LF exceeds that of GAMA15 at the $95\%$ level.

We fit a Schechter function \citep{schechter76} to our Ly$\alpha$
LF, where
\begin{equation}
\Phi(L)dL=\phi^{*}\left(\frac{L}{L^{*}}\right)^{\alpha}e^{-L/L^{*}}d\left(\frac{L}{L^{*}}\right).
\end{equation}

\noindent Our data lack the faint luminosity range needed to constrain
the faint-end slope, so we assume fixed values of $\alpha=-2.5$ and
$\alpha=-1.7$ that are meant to encompass the viable range
of high-redshift slopes. The $\alpha=-2.5$ value is consistent with
the best-fit values from NB Ly$\alpha$ surveys at $z=5.7$ and $z=6.6$
\citep{konno18,santos16}, while the $\alpha=-1.7$ value is consistent
with a deep spectroscopic survey at $z=5.7$ \citep{henry12}. Following
\citet{malhotra02}, we 
fit our LFs and estimate parameter errors using Cash Statistics \citep{cash79}.
We report the best-fit $\phi^{*}$ and $L^{*}$ values and errors
in Table \ref{lftable}.

\noindent 
\begin{figure*}
\includegraphics[bb=99bp 99.0291bp 712.8bp 544.66bp,clip,angle=180,width=8.5cm]{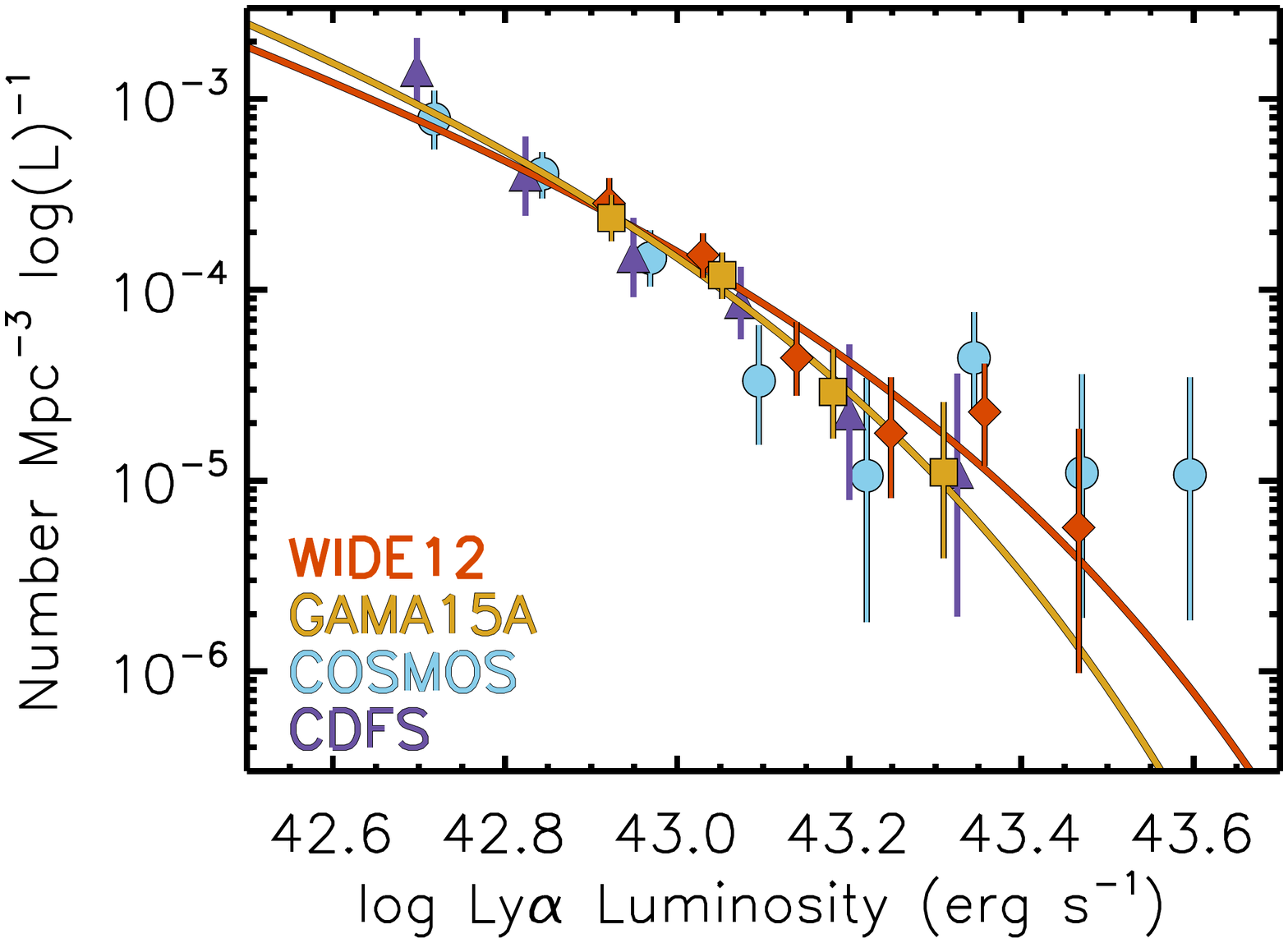}\includegraphics[bb=69.3bp 94.0777bp 712.8bp 534.757bp,clip,angle=180,width=8.5cm]{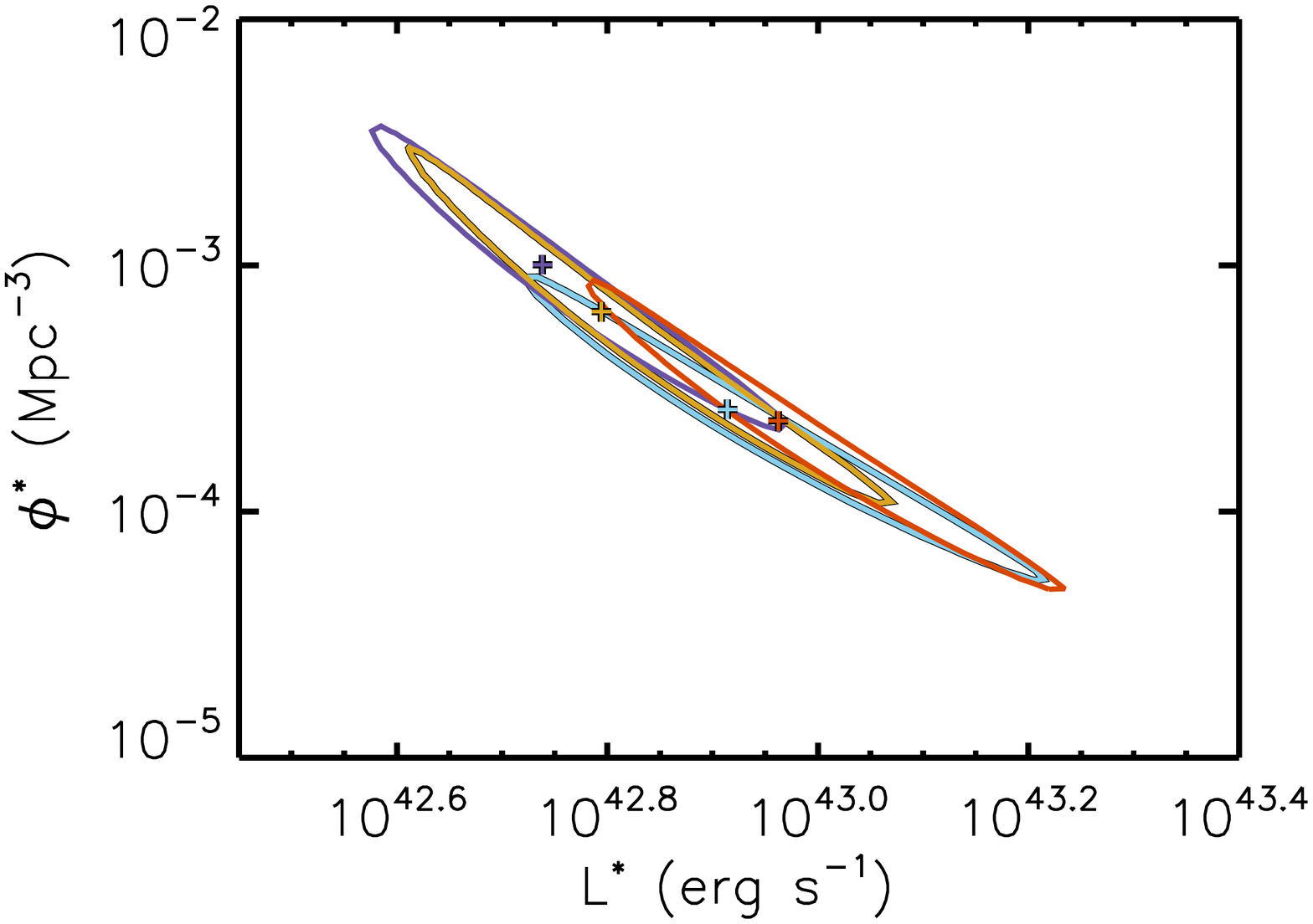}

\caption{(\textsl{Left}) Individual LAGER field Ly$\alpha$ LFs (WIDE12: orange diamonds; GAMA15A: yellow squares; COSMOS: blue
  circles; CDFS: purple triangles). The best-fit
Schechter functions are shown for fields WIDE12 and GAMA15A. (\textit{Right})
Corresponding 1$\sigma$ confidence intervals of the best-fit Schechter
parameters for all four LAGER fields. For the Schechter functions
and corresponding confidence contours, we assume a fixed faint-end
slope of $\alpha=-2.5$. We find good $1\sigma$ agreement between
all LAGER fields.}

\label{indiv_lf}
\end{figure*}

To assess the amount of Ly$\alpha$ light emitted relative to other
surveys we calculate the Ly$\alpha$ luminosity density,
\begin{equation}
\rho^{\text{{Ly}\ensuremath{\alpha}}}=\int L\Phi(L)dL.
\end{equation}

\noindent We adopt integration limits of $L_{{\rm {Ly}\alpha}}=10^{42.4}$
erg s$^{-1}$ to infinity which does not require large extrapolations
to unobserved faint luminosities and is consistent with previous NB
studies \citep{hu19,itoh18,konno18}. We list our luminosity density
values and 1$\sigma$ errors in Table \ref{lftable}. We compute 1$\sigma$
errors with a Monte-Carlo simulation that perturbs our LF data by
Poisson random deviates. This procedure is used to create ${\rm N}=\hbox{10,000}$
perturbed LFs. We then perform our standard LF fitting and luminosity
density calculation to find the inter-68-percentile of the resulting
luminosity density distribution.

\subsection{The 4-Field LAGER Ly$\alpha$ LF}

The areas covered by the four LAGER fields are $3.24$, $2.91$, $1.90$,
and $2.14$ deg$^{2}$ for WIDE12, GAMA15A, COSMOS, and CDFS, respectively.
The corresponding survey volumes for \lya\ galaxies at $z= 6.9$ are
$1.93$, $1.73$, $1.14$, and $1.29\times10^{6}$ Mpc$^{3}$, giving a
total survey volume of $6.1\times 10^{6} \hbox{Mpc}^{3}$. This is greater than
a factor of $\times2$ larger than other $z\sim7$ Ly$\alpha$ LFs
(see Table \ref{lfcomptable}). In Figure \ref{indiv_lf}, we show the WIDE12 and GAMA15A LAGER LFs
compared to the COSMOS and CDFS LAGER LFs presented in \citet{hu19}.
The LF best-fit parameters are found to agree within their $1\sigma$
Poisson errors, and we combine these data-sets to compute the 4-Field Ly$\alpha$ LF.

\begin{figure*}
\includegraphics[bb=99bp 99.0291bp 712.8bp 544.66bp,clip,angle=180,width=8.5cm]{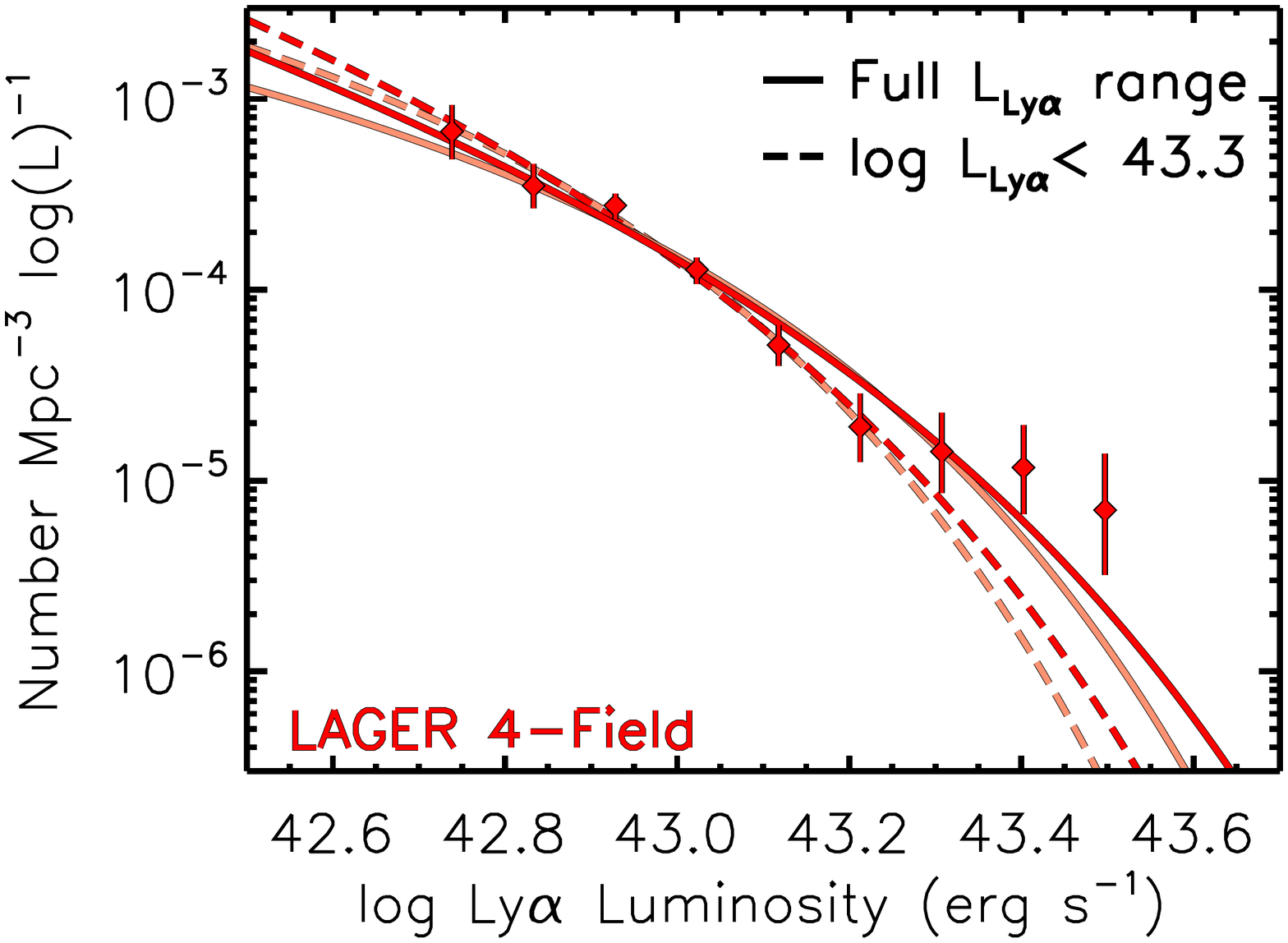}\includegraphics[bb=69.3bp 94.0777bp 712.8bp 534.757bp,clip,angle=180,width=8.5cm]{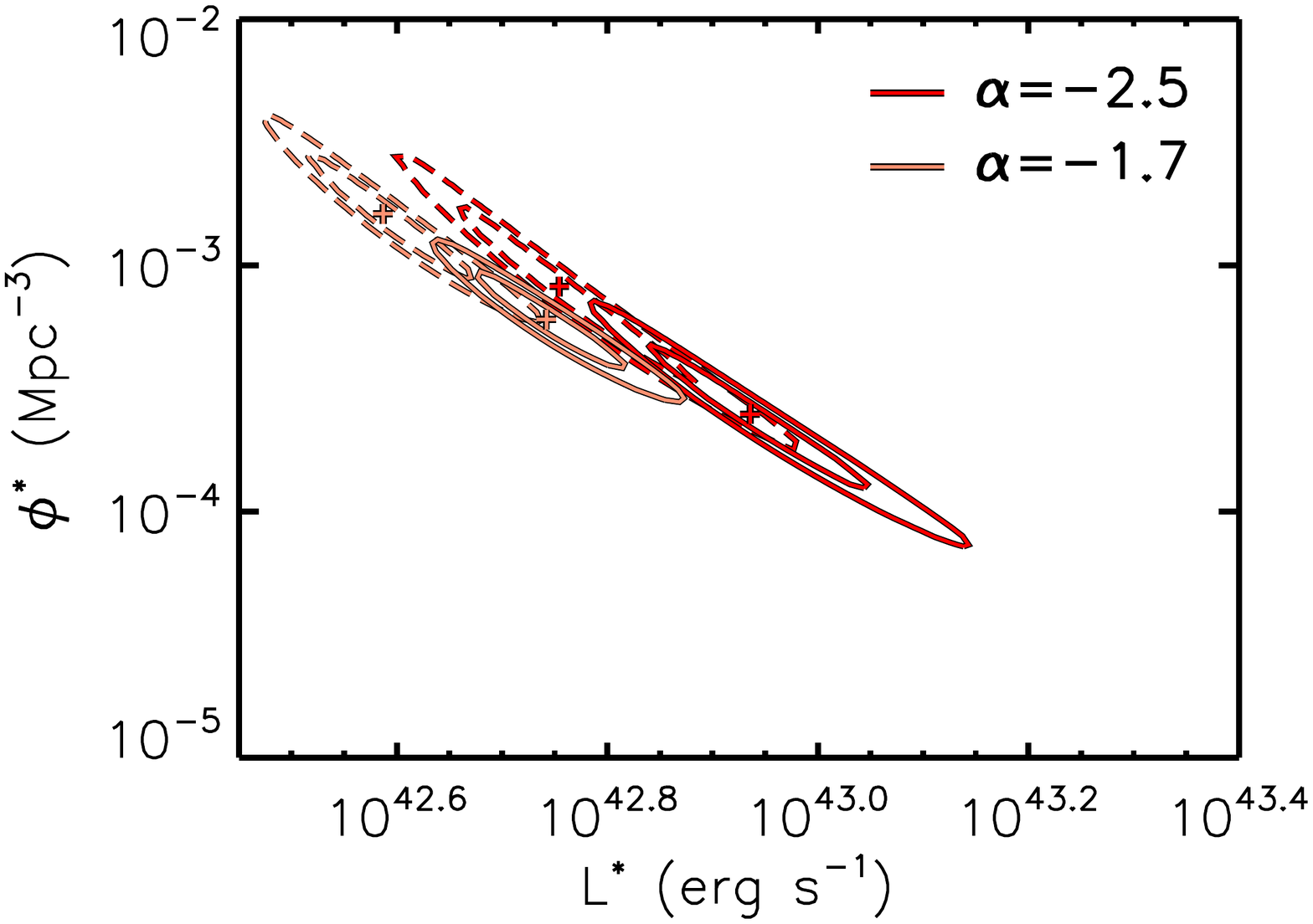}

\caption{(\textsl{Left}) The combined LAGER 4-Field Ly$\alpha$ LF and best-fit
Schechter functions. (\textit{Right}) Corresponding 1 and 2$\sigma$
confidence intervals for the best-fit Schechter parameters.}

\label{combo_lf}
\end{figure*}

In Figure \ref{combo_lf}, we show the 4-Field LAGER LF and our $1$
and $2\sigma$ confidence contours. We again see evidence in this
LF for a bright-end bump, and we illustrate the effect of excluding
the brightest luminosities data points ($L_{{\rm {Ly}\alpha}}>10^{43.3}$erg
s$^{-1}$) from our Schechter function fit in Figure \ref{combo_lf}.
We list all of our best-fit Ly$\alpha$ LF parameters in Table \ref{lftable}.
We find that our fit with a steep faint-end slope of $\alpha=-2.5$
is a marginally better description of our Ly$\alpha$ data points.
However, if the bright-end bump is excluded in the fits, then the
distinction between faint-end slopes largely goes away.

\section{Discussion}

\subsection{The Ly$\alpha$ IGM Transmission Fraction and the Neutral Hydrogen
Fraction}

\begin{figure*}
\begin{centering}
\includegraphics[bb=69.3bp 99.0291bp 712.8bp 544.66bp,clip,angle=180,width=15cm]{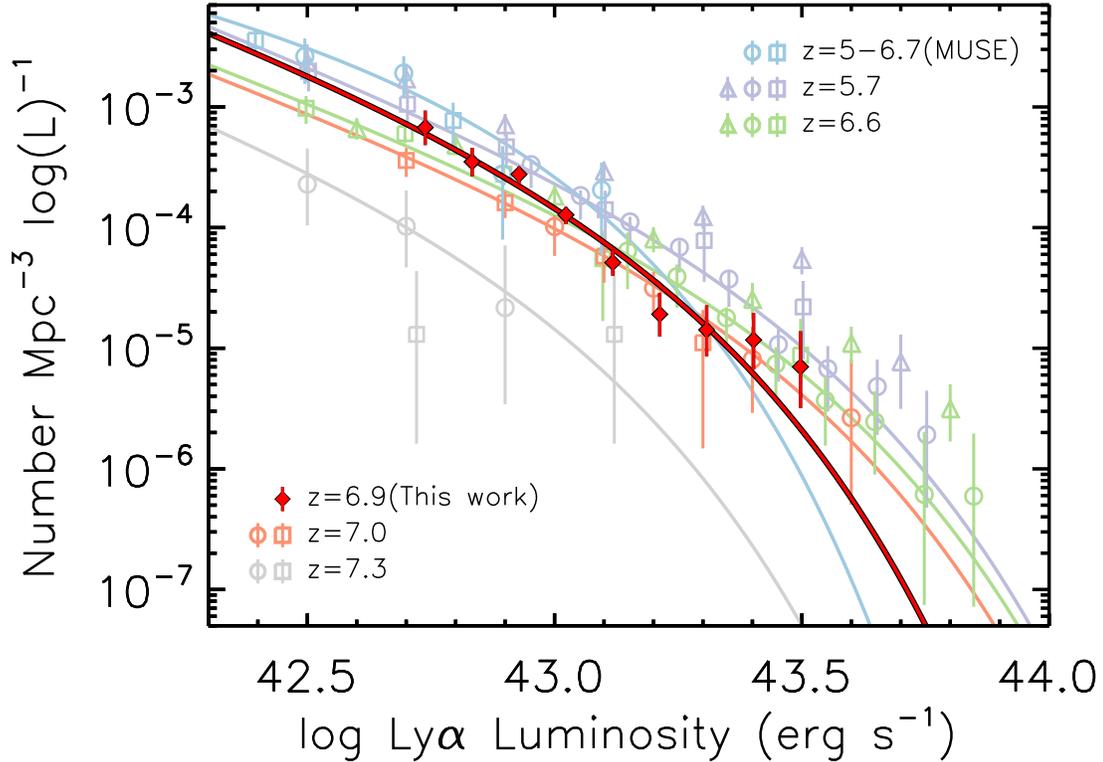}
\par\end{centering}
\caption{The combined $z=6.9$ LAGER 4-Field Ly$\alpha$ LF compared to archival
studies. We list the displayed archival references and corresponding point symbols
in Table \ref{lfcomptable}.
We show the best-fit $z=5.7$, $6.6$, and $7.3$ Schechter functions
with $\alpha\sim-2.5$ as compiled in \citet{itoh18}. We also show
the best-fit $3>z>6.7$ MUSE-Wide Ly$\alpha$ LF with $\alpha=-1.84$
from \citet{herenz19}. Within the luminosity range probed by MUSE-Wide,
this $3>z>6.7$ LF was found to be consistent with their highest redshift
Ly$\alpha$ LF at $5>z>6.7$ (shown by open blue circles).}
\label{comp_lf}
\end{figure*}

In Table \ref{lfcomptable} and Figure \ref{comp_lf}, we show how
our $z=6.9$ 4-Field LAGER Ly$\alpha$ LF compares to other high-redshift
LFs. We note that factor of $\sim2$ discrepancies between Ly$\alpha$
LFs at the same redshift are found between different NB surveys. While
some of this variation can be attributed to field-to-field variation,
survey systematics likely play a role \citep[e.g., see the discussion
in][]{taylor20}.

To measure the Ly$\alpha$ IGM transmission fraction at $z=6.9$,
we need to compare our Ly$\alpha$ LF to a reference LF that is at
a low enough redshift to be outside the reionization epoch, yet in
close redshift proximity to minimize any evolution of the LAE galaxy
population. We know from $z\sim6$ quasar spectra that the reionization
of the IGM is largely complete at $z\sim6$ \citep{Fan06}. Thus,
we compare our sample to the $z=5.7$ Ly$\alpha$ sample from \citet{konno18},
and we are particularly interested in accounting for any systematic
uncertainties between these surveys. Before computing the Ly$\alpha$ IGM
transmission fraction and neutral hydrogen fraction, we consider the
importance of systematics introduced by Ly$\alpha$ survey EW limits
and selection completeness.

\subsubsection{Ly$\alpha$ Survey EW Limits}
\label{sec:ewcompare}

\noindent \begin{deluxetable*}{ccccccc} 
\tablecolumns{5} 
\tablewidth{0pc} 
\tablecaption{Comparison of high-redshift LAE surveys.  The lower-limit EW$_{\rm{Ly}\alpha}^{\rm{lim}}$ are computed with the survey's NB excess criteria and corresponding NB and BB filters but normalized to a common assumed LAE spectrum and IGM attenuation prescription as described in section~\ref{sec:ewcompare}.}
\label{lfcomptable}
\tablehead{ 
\colhead{Reference} \vspace{-0.2cm} & \colhead{Redshift} & \colhead{Cosmic Age} & \colhead{Volume} & \colhead{EW$_{\rm{Ly}\alpha}^{\rm{lim}}$} & \colhead{log $L_{\rm{Ly}\alpha}^{\rm{lim}}$} & \colhead{Figure \ref{comp_lf}}\\
\colhead{ } & \colhead{ } & \colhead{(Gyr)} & \colhead{(Mpc$^{3}$)} & \colhead{(\AA)} & \colhead{log (erg s$^{-1}$)} & \colhead{symbol}}
\startdata 
This work & 6.93 & 0.76 & $6.1\times10^{6}$ & $10$ & $42.7$ & red diamond\\
\hline
\citet{herenz19} & 5.0-6.7 & 1.15-0.79 & $2.3\times10^{5}$ & --- & $42.2$ & blue circle\\
\citet{drake17} & 5.0-6.6 & 1.15-0.81 & $3.6\times10^{4}$ & --- & $41.0$ & blue square\\
\citet{santos16} & 5.7 & 0.98 & $6.3\times10^{6}$ & $6$ & $42.5$ & purple triangle\\
\citet{konno18} & 5.7 & 0.98 & $1.2\times10^{7}$ & $4$ & $42.9$ & purple circle\\
\citet{ouchi08} & 5.7 & 0.98 & $9.2\times10^{5}$ & $6$ & $42.5$ & purple square\\
\citet{matthee15} & 6.6 & 0.81 & $4.3\times10^{6}$ & $48$ & $42.5$ & green triangle\\
\citet{konno18} & 6.6 & 0.81 & $1.9\times10^{7}$ & $15$ & $43.0$ & green circle\\
\citet{ouchi10} & 6.6 & 0.81 & $8.0\times10^{5}$ & $48$ & $42.4$ & green square\\
\citet{itoh18} & 6.99 & 0.75 & $2.2\times10^{6}$ & $12$ & $42.9$ & red circle\\
\citet{ota17} & 7.02 & 0.75 & $6.1\times10^{5}$ & $0$ & $42.6$ & red square\\
\citet{konno14} & 7.30 & 0.71 & $2.5\times10^{5}$ & $0$ & $42.4$ & grey circle\\  
\citet{shibuya12} & 7.27 & 0.71 & $5.9\times10^{5}$ & $0$ & $42.7$ & grey square\\  
\enddata  
 \vspace{-0.9cm}
\end{deluxetable*}

We show in Table \ref{lfcomptable} that archival surveys have adopted
different EW limits when computing their Ly$\alpha$ LFs.
The EW limits listed in Table
  \ref{lfcomptable} have been re-computed in a consistent manner so
  that relative comparisons can be made. Following the procedure we used for our LAGER Ly$\alpha$
  flux measurements (see Section \ref{select}),  we assume the LAE
  candidates have a delta-function Ly$\alpha$ line profile at the
  center of the NB filter and a UV continuum slope of $-2$ that is
  attenuated by the IGM via the \citet{inoue14} model. Given this
  assumed spectral shape, the NB/BB filter responses, and the
  study-specific NB-excess cut, we compute the survey's EW limit.

To estimate the importance of the EW selection difference, we
consider a non-evolving EW distribution parameterized by a declining
exponential and a scale length of $100$ \AA. Given the extremes
of the EW lower-limits shown in Table \ref{lfcomptable}, we estimate
that the different EW limits could at most account for a factor of
$\sim1.6$ number density variations. In general, we find that the
survey-to-survey variation introduced by this variation should be
relatively small with the caveat that we have not considered an evolving EW 
scale length with redshift or other scale length dependencies. Using
the same methodology, we specifically compare our EW $>10$ \AA~
survey at $z=6.9$ to Konno et al.'s EW $>4$ \AA~ survey at $z=5.7$.
We find that if their survey adopted an EW $>10$ \AA~ cut, their
Ly$\alpha$ number density would decline by a small factor of $\sim1.06$.

\subsubsection{Selection Completeness}

Recently, another source of systematic error was emphasized by \citet{hu19},
selection completeness. They found that high-redshift NB surveys can
lose a significant fraction of their survey area by requiring non-detections
in all of their veto bandpasses. However, this effect is not accounted
for in the high-redshift NB surveys that we compare to and show in
Figure \ref{comp_lf}. The size of the area lost to this effect depends
on several variables including veto band depth, seeing, extraction
aperture size, and the scope of the applied bright star mask, and
it is unlikely to be constant between surveys. For example, \citet{hu19}
found that about $40\%$ of their survey area is masked by $3\sigma$
veto sources. While for our WIDE12 and GAMA15A fields with shallower
veto bandpasses, we find that only $\sim15\%$ of our survey area
is masked. By accounting for this incompleteness, we compute accurate
survey areas and find luminosity densities that are consistent within
$1\sigma$ between our combined WIDE12+GAMA15A and the combined COSMOS+CDFS
measurements.

The $z=5.7$ NB Ly$\alpha$ LFs do not account for selection incompleteness,
so we quantify this effect, which - if left unaccounted for - will
cause an over-estimate of the survey's area and a corresponding under-estimate
of the LF's normalization, $\phi^{*}$ \citep[see][for details]{hu19}.
Konno et al.\ use the deeper $z=5.7$ Ly$\alpha$ survey from \citet{ouchi08}
to constrain the faint-end of their LF fit. Their resulting combined
fit to these two different data-sets with different veto-band depths
makes it difficult to estimate an effective selection incompleteness.
To simplify the problem, we re-fit Konno et al.'s $z=5.7$ Ly$\alpha$
LF assuming a fixed faint-end slope of $\alpha=-2.5$ but without
the Ouchi et al.\ data.  We compute a Ly$\alpha$ luminosity density
of $3.70\times10^{39}$ erg s$^{-1}$Mpc$^{-3}$ which is within $1\sigma$
of the published $z=5.7$ result \citep[$3.49^{+0.58}_{-0.71}\times10^{39}$erg s$^{-1}$Mpc$^{-3}$][]{konno18,itoh18}.

Konno et al.\ use HSC-SSP internal data release S16A $g$-band images
to construct their $g>g_{3\sigma}$ veto-band selection over four
fields: COSMOS, SXDS, DEEP2-3, and ELAIS-N1. The publicly available
dataset with the closest image characteristics to S16A is HSC-SSP
DR1, and we use a $0.5\times0.5$ deg$^{2}$ HSC-SSP DR1 COSMOS $g$-band
image in the UltraDeep layer to characterize their veto-band selection
completeness. We randomly place ${\rm N}=\hbox{10,000}$ $D=1.5''$ apertures in
our DR1 $g$-band sub-image and find that $16\%$ of these random
apertures would not be selected due to overlap with $3\sigma$ isophotes
of foreground objects. However, Konno et al.\ apply a bright star
mask that removes $\sim7\%$ of the survey area within our sub-field
\citep{coupon18}, implying a final selection incompleteness of $\sim9\%$.
Correcting for this estimated selection incompleteness and the different
EW limits, we compute a $z=5.7$ Ly$\alpha$ luminosity density of
$3.84_{-0.71}^{+0.64}\times10^{39}$ erg s$^{-1}$Mpc$^{-3}$. The
overall correction that we apply is small and simply using Konno et
al.'s published luminosity density value would not significantly
alter our results.

\begin{deluxetable*}{cccc}
\tablecolumns{3} 
\tablewidth{0pc} 
\tablecaption{IGM neutral hydrogen fraction constraints.}
\label{neutraltable}
\tablehead{ 
\colhead{Reference and } \vspace{-0.2cm} &  \colhead{Observational} & \colhead{Model $x_{\rm{HI}}$}\\
\colhead{Relevant Figures}   & \colhead{Constraints} & \colhead{at $z=6.9$}}
\startdata 
\citet{malhotra06} Fig.1 \& 2 & $T^{\rm{IGM}}$ \& Ly$\alpha$ number density & $0.00 - 0.04$ \\
\citet{dijkstra07} Fig.6 \& \citet{furlanetto06} Fig.1 & $T^{\rm{IGM}}$ & $0.00 - 0.16$ \\
\citet{mcquinn07} Fig.5 & cumulative Ly$\alpha$ LF & $0.00 - 0.33$ \\
\enddata  
 \vspace{-0.9cm}
\end{deluxetable*}

\subsubsection{Ly$\alpha$ IGM Transmission Fraction\label{t_igm}}

We compute the Ly$\alpha$ IGM transmission fraction via:
\begin{equation}
\frac{T_{z=6.9}^{{\rm {IGM}}}}{T_{z=5.7}^{{\rm {IGM}}}}=\frac{\rho_{z=6.9}^{{\rm {Ly}\alpha}}/\rho_{z=5.7}^{{\rm {Ly}\alpha}}}{\rho_{z=6.9}^{{\rm {UV}}}/\rho_{z=5.7}^{{\rm {UV}}}}= 
1.0_{-0.28}^{+0.26},
\end{equation}

\noindent which assumes unchanging ISM properties and stellar populations
from $z=5.7$ to $6.9$, a $200$ Myr duration \citep[see also][for a similar procedure]{zheng17,konno18,itoh18,hu19}.
We use the corrected $z=5.7$ Ly$\alpha$ luminosity density of $3.84\times10^{39}$
erg s$^{-1}$Mpc$^{-3}$ and our $z=6.9$ 4-Field LAGER Ly$\alpha$
luminosity density of $2.44\times10^{39}$ erg s$^{-1}$Mpc$^{-3}$.
These luminosity densities are computed with consistent
integration limits and a fixed faint-end slope of $\alpha=-2.5$.
We compute a Ly$\alpha$ luminosity density ratio of $\rho_{z=6.9}^{{\rm {Ly}\alpha}}/\rho_{z=5.7}^{{\rm {Ly}\alpha}}=0.63_{-0.15}^{+0.13}$
and use the UV ratio of $\rho_{z=6.9}^{{\rm {UV}}}/\rho_{z=5.7}^{{\rm {UV}}}=0.63\pm0.09$
from \citet{finkelstein15}. These ratios are consistent with the
UV and Ly$\alpha$ LFs evolving at a similar rate from $z=5.7$ to
$6.9$ and imply a Ly$\alpha$ IGM transmission fraction of 
$T_{z=6.9}^{{\rm {IGM}}}/T_{z=5.7}^{{\rm {IGM}}}=1.0_{-0.28}^{+0.26}$.

\begin{figure}
\begin{centering}
\includegraphics[clip,width=8.5cm]{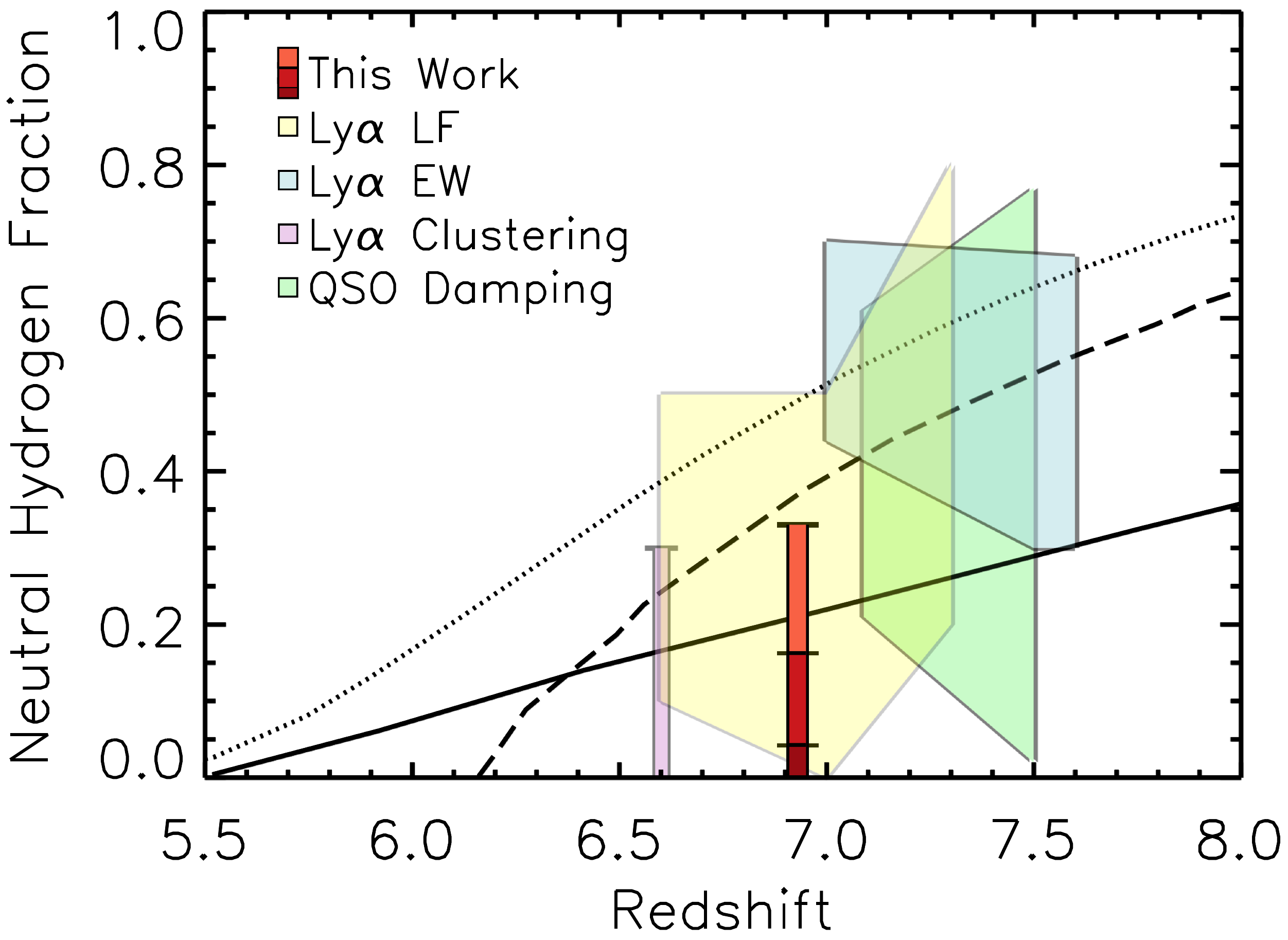}
\par\end{centering}
\caption{Evolution of the volume-averaged $x_{HI}$.
Our $1\sigma$ upper limit results based on three
  different Ly$\alpha$ transmission models (red
shaded bars, see Table \ref{neutraltable} and Section \ref{x_igm} for details) are shown relative to other observational
constraints. While we adopt the most conservative limit of $x_{HI}<0.33$
at $z=6.9$, we show the more restrictive constraints favored by
alternative reionization models with different
shading. 
Our observational constraint favors the more gradual reionization
scenario proposed by \citet[][solid curve]{finkelstein19} while being at odds with
the late and rapid reionization scenario presented by \citet[][dashed
curve]{robertson15} and \citet[][dotted curve]{kulkarni19a}. The displayed regions show the parameter space allowed
by the $\pm1\sigma$ results from archival studies.  The Ly$\alpha$ LF
region is defined by the results from \citet{konno18}, \citet{itoh18},
\citet{konno14}, and \citet{inoue18}. The Ly$\alpha$ EW
region is defined by the results from \citet{mason18}, \citet{tilvi14}, and
\citet{jung20}. The Ly$\alpha$
clustering region is defined by the results from \citet{ouchi18}. The
QSO damping region is defined by the results from \citet{greig17},
\citet{greig19}, and \citet{banados18}.}

\label{neu_frac}
\end{figure}

\subsubsection{Volume-Averaged Neutral Hydrogen Fraction \label{x_igm}}

The conversion of Ly$\alpha$ transmission fractions
to a neutral hydrogen fraction $x_{HI}$
is model dependent. By employing multiple theoretical models, we quantify
systematic errors. All of the following $x_{HI}$
estimates are volume-averaged neutral hydrogen fractions.

First, we use our Ly$\alpha$ LF to constrain the radiative transfer
simulations of \citet{mcquinn07} which display the expected drop
in the cumulative Ly$\alpha$ LF in their Figure 5. 
Taking the $1\sigma$ lower bound on transmission ($T_{-1\sigma} = 0.72$),
this figure implies upper limits 
of $x_{HI}<0.16$, 
 $x_{HI}<0.33$, and
  $x_{HI}<0.11$, respectively,
for \citet{mcquinn07}'s models I, II, and III.    The models differ in their assumed scaling of ionizing 
photon production with halo mass.   Model I is a straight proportionality, while model II gives greater
weight to the most massive halos ($\dot{N}_{\rm ion} \propto M^{5/3}$), resulting in larger bubbles and
corresponding qualitatively to the recent ``oligarchs'' model \citep{naidu20}.
Model III has yet smaller bubbles due to minihalo absorption of Lyman continuum photons.

We also estimate the $z=6.9$ neutral hydrogen fraction with a Ly$\alpha$
volume test where each observed LAE is taken as evidence for an ionized
bubble allowing for Ly$\alpha$ escape \citep{malhotra06}. Using
$T_{z=6.9}^{{\rm {IGM}}}/T_{z=5.7}^{{\rm {IGM}}}=1.0_{-0.28}^{+0.26}$,
a Ly$\alpha$ velocity shift of $200-300$ km s$^{-1}$ \citep[e.g.,][]{mclinden11},
a correlation length of $4$ Mpc, and a Ly$\alpha$ number density
of $\sim1.5\times10^{-4}$ Mpc$^{-3}$, we find $x_{HI}<0.04$
via \citet[][Figure 1 and 2]{malhotra06}. 

Finally, we use the analytical models of \citet{dijkstra07} and \citet{furlanetto06}
to estimate the typical size of ionized bubbles and then use this
value to constrain the neutral hydrogen fraction to be $x_{HI}<0.16$.
We summarize all of the $x_{HI}$
constraints in Table \ref{neutraltable}. 

Adopting a conservative limit, we conclude that $x_{HI}<0.33$
at $z=6.9$.  In Figure \ref{neu_frac}, we show our $z=6.9$ constraint
in comparison to model predictions and other observational constraints.
Our $x_{HI}$
estimate is consistent with other results based on Ly$\alpha$ LFs
\citep{konno14,konno18,itoh18,inoue18}, Ly$\alpha$ clustering \citep{ouchi18},
Ly$\alpha$ EW tests (\citealt{mason18,tilvi14,jung20}), and QSO damping wings \citep{greig17,banados18,greig19}. Our
observational constraint favors the more gradual reionization
scenario proposed by \citet{finkelstein19} but is in tension with a 
rapid reionization scenario preferred by \citet{robertson15} and
\citet{kulkarni19a}.

With the conclusion of LAGER we will more
than double our current survey area enabling us to further constrain
$x_{HI}$
based on Ly$\alpha$ LF tests. Additionally, the increased Ly$\alpha$
sample size will allow us to conduct Ly$\alpha$ clustering analyses
which will give us another handle on the $z=6.9$ hydrogen neutral
fraction.

\section{Summary}

We present the LAGER 4-Field $z=6.9$ Ly$\alpha$ luminosity function.
We identified $N=95$ new Ly$\alpha$ emitters selected in the WIDE12
and GAMA15A fields and combine this sample with the existing COSMOS
and CDFS LAGER Ly$\alpha$ sample. We compute the 4-Field LAGER Ly$\alpha$
luminosity function with a survey volume of $6.1\times10^{6}$ Mpc$^{3}$.
This is more than double the size of the next largest $z\sim7$ Ly$\alpha$
survey. We investigate potential systematics in NB Ly$\alpha$ surveys
and taking these issues into account we estimate a Ly$\alpha$ IGM
transmission of $T_{z=6.9}^{{\rm {IGM}}}/T_{z=5.7}^{{\rm {IGM}}}=1.0_{-0.28}^{+0.26}$
which is consistent with a fully ionized $z=6.9$ neutral hydrogen
fraction.  Using our $1\sigma$ lower bound on transmission
($T_{-1\sigma} = 0.72$), we find that the most conservative
reionization model predicts a volume-averaged neutral hydrogen fraction of  $x_{HI} < 0.33$.

\noindent \acknowledgments{We thank the anonymous referee for comments that
substantially improved the manuscript. IGBW
is supported by an appointment to the NASA Postdoctoral Program
at the Goddard Space Flight Center, administered by the Universities
Space Research Association through a contract with NASA. The material is based upon work supported by NASA under award number 80GSFC21M0002. The
LAGER survey has been supported in part by the US National
Science Foundation through NSF grant AST-1518057.

This work was funded by ANID projects CATA-BASAL AFB-170002 (L.F.B., F.E.B.), FONDECYT Regular 1190818 (F.E.B.) and 1200495 (F.E.B.), and Millennium Science Initiative Program ICN12\_009 (F.E.B.).

J.X.W. and W.D.H are supported by National Science Foundation of China
(grants No. 11421303 \& 11890693) and CAS Frontier Science Key
Research Program (QYZDJ-SSW-SLH006). Z.Y.Z. thanks the National
Science Foundation of China (11773051, 12022303) and the CAS Pioneer
Hundred Talents Program. 
}

\acknowledgments{
This project used data obtained with the Dark Energy Camera (DECam), which was constructed by the Dark Energy Survey (DES) collaboration. Funding for the DES Projects has been provided by the U.S. Department of Energy, the U.S. National Science Foundation, the Ministry of Science and Education of Spain, the Science and Technology Facilities Council of the United Kingdom, the Higher Education Funding Council for England, the National Center for Supercomputing Applications at the University of Illinois at Urbana-Champaign, the Kavli Institute of Cosmological Physics at the University of Chicago, Center for Cosmology and Astro-Particle Physics at the Ohio State University, the Mitchell Institute for Fundamental Physics and Astronomy at Texas A\&M University, Financiadora de Estudos e Projetos, Funda\c c\~ao Carlos Chagas Filho de Amparo, Financiadora de Estudos e Projetos, Funda\c c\~ao Carlos Chagas Filho de Amparo \`a Pesquisa do Estado do Rio de Janeiro, Conselho Nacional de Desenvolvimento Cient\'ifico e Tecnol\'ogico and the Minist\'erio da Ci\^encia, Tecnologia e Inova\c c\~ao, the Deutsche Forschungsgemeinschaft and the Collaborating Institutions in the Dark Energy Survey.

The Collaborating Institutions are Argonne National Laboratory, the University of California at Santa Cruz, the University of Cambridge, Centro de Investigaciones En\'ergeticas, Medioambientales y Tecnol\'ogicas-Madrid, the University of Chicago, University College London, the DES-Brazil Consortium, the University of Edinburgh, the Eidgen\"ossische Technische Hochschule (ETH) Z\"urich, Fermi National Accelerator Laboratory, the University of Illinois at Urbana-Champaign, the Institut de Ci\`encies de l'Espai (IEEC/CSIC), the Institut de F\'isica d'Altes Energies, Lawrence Berkeley National Laboratory, the Ludwig-Maximilians Universit\"at M\"unchen and the associated Excellence Cluster Universe, the University of Michigan, the National Optical Astronomy Observatory, the University of Nottingham, the Ohio State University, the OzDES Membership Consortium, the University of Pennsylvania, the University of Portsmouth, SLAC National Accelerator Laboratory, Stanford University, the University of Sussex, and Texas A\&M University. 

Based on observations at Cerro Tololo Inter-American Observatory, NSF--NOIRLab (NOIRLab Prop. ID 2017A-0366, 2018A-0371, 2018B-0327; PI: S. Malhotra; 2018B-0907, 2019A-0912; PI: L. F. Barrientos; 2017A-0920; PI: L. Infante), which is operated by the Association of Universities for Research in Astronomy (AURA) under a cooperative agreement with the National Science Foundation.
}

\acknowledgments{
The Hyper Suprime-Cam (HSC) collaboration includes the astronomical communities of Japan and Taiwan, and Princeton University. The HSC instrumentation and software were developed by the National Astronomical Observatory of Japan (NAOJ), the Kavli Institute for the Physics and Mathematics of the Universe (Kavli IPMU), the University of Tokyo, the High Energy Accelerator Research Organization (KEK), the Academia Sinica Institute for Astronomy and Astrophysics in Taiwan (ASIAA), and Princeton University. Funding was contributed by the FIRST program from Japanese Cabinet Office, the Ministry of Education, Culture, Sports, Science and Technology (MEXT), the Japan Society for the Promotion of Science (JSPS), Japan Science and Technology Agency (JST), the Toray Science Foundation, NAOJ, Kavli IPMU, KEK, ASIAA, and Princeton University.

This paper makes use of software developed for the Large Synoptic Survey Telescope. We thank the LSST Project for making their code available as free software at http://dm.lsst.org

This paper is based in part on data collected at the Subaru Telescope
and retrieved from the HSC data archive system, which is operated by
Subaru Telescope and Astronomy Data Center at National Astronomical
Observatory of Japan. Data analysis was in part carried out with the
cooperation of Center for Computational Astrophysics, National
Astronomical Observatory of Japan.

This publication makes use of data products from the Two Micron All
Sky Survey, which is a joint project of the University of
Massachusetts and the Infrared Processing and Analysis
Center/California Institute of Technology, funded by the National
Aeronautics and Space Administration and the National Science
Foundation.

The Pan-STARRS1 Surveys (PS1) and the PS1 public science archive have been made possible through contributions by the Institute for Astronomy, the University of Hawaii, the Pan-STARRS Project Office, the Max-Planck Society and its participating institutes, the Max Planck Institute for Astronomy, Heidelberg and the Max Planck Institute for Extraterrestrial Physics, Garching, The Johns Hopkins University, Durham University, the University of Edinburgh, the Queen\textquoteright s University Belfast, the Harvard-Smithsonian Center for Astrophysics, the Las Cumbres Observatory Global Telescope Network Incorporated, the National Central University of Taiwan, the Space Telescope Science Institute, the National Aeronautics and Space Administration under Grant No. NNX08AR22G issued through the Planetary Science Division of the NASA Science Mission Directorate, the National Science Foundation Grant No. AST-1238877, the University of Maryland, Eotvos Lorand University (ELTE), the Los Alamos National Laboratory, and the Gordon and Betty Moore Foundation.
}

\bibliographystyle{aasjournal} 
\bibliography{wold}

\end{document}